\documentclass[12pt]{JHEP3}

\def\ee{e^+e^-}
\def\qq{q\bar q}
\def\qqg{q\bar qg}
\def\HW{\textsc{\small HERWIG}}
\def\pythia{\textsc{\small PYTHIA}}
\def\hpp{\mbox{\textsf{Herwig++}}}
\def\ThePEG{\textsf{ThePEG}}
\def\Pythia7{\textsf{Pythia7}}
\def\SHERPA{\textsf{SHERPA}}
\def\CLHEP{\textsf{CLHEP}}
\def\ycut{y_{\rm cut}}

\def\Clm{{\texttt{Cl}_{\texttt{max}}}}
\def\Clp{{\texttt{Cl}_{\texttt{pow}}}}
\def\D{\cite{DELPHIshapes}}
\def\O{\cite{OPALnjet}}
\def\DF{\cite{DELPHIfourj}}
\def\OC{\cite{Acton:1991aa}}
\def\Sx{\cite{SLD03}}
\def\A{\cite{ALEPHLambda}}
\def\SB{\cite{SLDbfrag}}
\def\AB{\cite{ALEPHbfrag}}

\usepackage{epsfig, bm, amsmath}

\title{\hpp{} 1.0:\\An Event Generator for {\boldmath
    $e^+e^-$} Annihilation} 
\author{Stefan Gieseke$^{\dagger}$,
  Alberto Ribon$^{\ddagger}$, Michael H Seymour$^{\star}$, Philip
  Stephens$^{\dagger}$,
  Bryan~Webber$^{\dagger,\S}$\\
  $^{\dagger}$Cavendish Laboratory, University of Cambridge,
  Madingley Road, Cambridge, CB3~0HE, United Kingdom.\\
  $^{\ddagger}$EP Division, CERN, 1211 Geneva 23, Switzerland.\\
  $^{\star}$Theory Group, Department of Physics and Astronomy,
    Schuster Laboratory, University of Manchester, Manchester M13 9PL,
    UK.\\
  $^{\S}$TH Division, CERN, 1211 Geneva 23, Switzerland.}

\abstract{Results from the new Monte Carlo event generator \hpp{}
  are presented. This first version simulates Hadron
  Emission Reactions With Interfering Gluons in electron--positron
  annihilation.  The parton shower evolution is carried out
  using new evolution variables suited to describing radiation from
  heavy quarks as well as light partons.
  The partonic final state is fragmented into hadrons by means of an
  improved cluster hadronization model. The results are compared with
  a wide variety of data from LEP and SLC.}


\preprint{Cavendish-HEP-03/19\\CERN-TH/2003-265}

\begin{document}
\section{Introduction}
The new generation of high energy colliders such as the Large Hadron
Collider (LHC) or a future linear collider (NLC) require new tools for
the simulation of signals and backgrounds.  The widely used event generators
\HW{} \cite{Herwig64} and \pythia{} \cite{Pythia} underwent tremendous
development during the LEP era and have reached the limit of
reasonable maintenance in the future.  Therefore these programs
(\Pythia7{}) \cite{Pythia7} as well as new projects, like \SHERPA{}
\cite{SHERPA}, are being completely (re-)developed in the object-oriented
programming language C++.

In this paper we present results from the new Monte Carlo event
generator \hpp\ as the first step in the redevelopment of \HW{}.  The
generator will be used here to simulate $\ee$ annihilation events.  In
order to have full control of the basic physics steps that are
simulated, we need to put the new generator on a firm basis with
respect to LEP and SLC results before we go on to upgrade it to
initial-state showers and the other requirements for the simulation of
lepton-hadron and hadron-hadron collisions.  Therefore we have tested
the predictions of the generator against a wide range of observables
that have been measured at LEP and SLC, and have explored the
sensitivity to the most important parameters and cutoffs.  We did not
perform a high-precision tuning: our aim here is rather to describe
the program and to show that it is able to give results as acceptable
as those generated by its predecessor \HW{} for a reasonable choice of
parameters.

\section{Main features of the code}
The details of \hpp{} will be described in conjunction with
the release of the code~\cite{manual}.
The main stages of the simulation of
$\ee$ annihilation are the same as in \HW{} \cite{Herwig64}.
However, in comparison to its predecessor, \hpp{} features a new parton
shower and an improved cluster hadronization model.  At present,
hadronic decays are implemented in the same fashion as they were in
\HW{}. 

The program is based on the Toolkit for High Energy Physics Event
Generation (\ThePEG{}) \cite{ThePEG} and the Class Library for High
Energy Physics (\CLHEP{}) \cite{CLHEP}.  They are utilized in order to
take advantage of the extended general functionality they can provide.
The usage of \ThePEG{} unifies the event generation framework with
that of \Pythia7{}.  This will provide benefits for the user, as the
user interface, event storage etc.\ will appear to be the same.  The
implementations of the physics models, however, are completely
different and independent from each other.

Our simulation starts with an initial hard process $\ee\to(\gamma^*,
Z^0)\to\qq+\gamma\gamma$.  The final state photons simulate QED
radiation from the initial state, so that a radiative return can be
properly simulated.  For the present paper we shall only be interested
in the details of the QCD parton shower in the final state.  The
final-state parton shower starts with a quark and antiquark that carry
momenta $p_q$ and $p_{\bar q}$, respectively, and have an invariant
mass squared of $Q^2=(p_q+p_{\bar q})^2$.  The only detail we are
concerned with in relation to initial-state radiation is that the
centre-of-mass frame of the $\qq$--pair is slightly boosted with
respect to the collider laboratory frame and that $Q$ may be different
from the $\ee$ centre-of-mass energy.  We have made sure that the
applied cuts on the energy of the annihilating $\ee$ subsystem are the
same as those used in the experimental analyses.

\subsection{Parton shower} The partonic evolution from the large scale
of the hard collision process down to hadronic scales via the coherent
emission of partons, mainly gluons, is simulated on the basis of the
Sudakov form factor.  Starting from the hard process scale $Q_0$,
subsequent emissions at scales $Q_i$ and momentum fractions $z_i$
are randomly generated as a Markov
chain on the basis of the soft and collinear approximation to partonic
matrix elements.  Details are described in chapter 5 of \cite{book}.
In \hpp{} we have chosen a new framework of variables, generically
called $(\tilde q, z)$.  Here, $\tilde q$ is a scale that appears
naturally in the collinear approximation of massive partonic matrix
elements and generalizes the evolution variable of \HW{} to the
evolution of massive quarks.  $z$ is a relative momentum
fraction; the evolution is carried out in terms of the Sudakov
decomposition of momenta in the frame where the respective colour
partners are back-to-back.  As in \HW, the use of the new variables
allows for an inherent angular ordering of the parton cascade, which
simulates coherence effects in soft gluon emission.  The details of
the underlying formalism are described elsewhere~\cite{NewVariables}.

The most important parameter of the parton shower that we will be
concerned with in this paper is the cutoff parameter $Q_g$, which
regularizes the soft gluon singularity in the splitting functions and
determines the termination of the parton shower.  Less important but
relevant in extreme cases is the treatment of the strong coupling
constant at low scales.  We have parametrized $\alpha_S(Q)$ below a
small scale $Q_{\rm min} > \Lambda_{QCD}$ in different ways.  We keep
$Q_{\rm min}$ generally to be of the order of 1\,GeV, where we expect
non-perturbative effects to become relevant.  Below that scale
$\alpha_S (Q)$ can optionally be
\begin{itemize}
\item set to zero, $\alpha_S(Q<Q_{\rm min}) = 0$, 
\item frozen, $\alpha_S(Q<Q_{\rm min}) = \alpha_S(Q_{\rm min})$, 
\item linearly interpolated in $Q$, between 0 and $\alpha_S(Q_{\rm
    min})$,
\item quadratically interpolated in $Q$, between 0 and
  $\alpha_S(Q_{\rm min})$\,.
\end{itemize}
We put the final partons of the shower evolution on their constituent
mass shells, since the non-perturbative cluster hadronization will
take over at this scale, so we usually have kinematical constraints
that keep $Q$ above $Q_{\rm min}$, in which case the treatment below
$Q_{\rm min}$ is irrelevant.  Typically, $\alpha_S(Q_{\rm min}) \sim
1$ here.

\subsection{Hadronization and decay} 
The partonic final state is turned into a hadronic final state within
the framework of the cluster hadronization model of \HW\ 
\cite{Webber:1983if}.  In order to address some
shortcomings~\cite{kupco} a new cluster hadronization
model~\footnote{A new cluster hadronization model 
that addresses some of these shortcomings
is also a feature of \SHERPA~\cite{Winter:2003tt}.} has been
created for \hpp{}, which is discussed in sec.~\ref{sec:hadro}.  The
emerging hadrons are possibly unstable and eventually decay.  The
decay matrix elements and modes correspond to those in \HW{}.

\section{The parton shower in detail}
\label{sec:partonshower}

\subsection{Hard matrix element correction}
\label{sec:hardmec}
Before we begin the parton shower evolution, but after obtaining the
final state $\qq$--pair from the hard process, we decide whether or
not a so-called hard matrix element correction will be applied.  In
order to do so, we decompose the $\qqg$--phase space into regions that
will be covered by the parton shower emissions and a `dead' region
that, based on our choice of evolution variables and initial
conditions, is never populated by first parton shower emissions (see
\cite{NewVariables}).  To take into account gluon emissions into the
dead region we generate a pair of three--body phase-space variables
$x, \bar x$ according to the first order QCD matrix element.  However,
we only accept emissions into the dead region of phase space at a rate
that is given by the QCD matrix element, that is, only 3 \% of all
emissions are corrected by the hard matrix element at all. Once we
accept an additional hard gluon emission, we replace the $\qq$--final
state with the $\qqg$ final state.  We keep the orientation of either
the quark or antiquark with weights $x^2$ and $\bar x^2$ respectively,
resulting in properly oriented three-jet events apart from finite mass
effects~\cite{Kleiss:1986re}.  In this way, we take into account the
most important subleading higher-order corrections that are not
enhanced by additional soft or collinear logarithms.

\subsection{Initial conditions}
\label{sec:initialcond}
Having completed the hard matrix element correction, the next task is
to determine the initial conditions for the parton shower evolution.
For every particle $a$ we determine the colour partner or, more
generally, the gauge `charge' partner $\bar a$.  In the case of a
$\qq$ final state there is no ambiguity, whereas the gluon in $\qqg$
is assigned the quark or the antiquark with equal probability.

For different interactions there can be different `charge' partners.
In our case we have also implemented collinearly enhanced photon
emission from charged particles.  In the case of the $\qqg$ final
state the gluon doesn't radiate photons and the only two charge
partners are the quark and the antiquark.  The remaining parts of the
shower evolution are carried out in exactly the same way.  Different
sorts of interaction just add another splitting possibility for a given
particle, which will compete with the others for the next possible
splitting that occurs.   

Once the colour partners are determined, we fix the shower kinematics
and the initial evolution scale.  As explained in detail in
\cite{NewVariables}, the shower evolution of a particle $a$ is carried
out in a Sudakov basis, 
\begin{equation}
  \label{eq:sudakovbasis}
  q=\alpha p_a + \beta n_{\bar a} + q_\perp,
\end{equation}
where $p_a$ is the momentum of particle $a$ with (current) mass-squared
$p_a^2 = m_a^2$,  $n_{\bar a}$ is a lightlike vector in the `backwards'
direction, along the momentum $p_{\bar a}$ of the partner $\bar a$,
and $q_\perp$ is the transverse
momentum, $q_\perp\cdot p_a = q_\perp\cdot n_{\bar a} = 0$.
In the centre-of-mass frame of $a$ and $\bar a$ we have
$p_a=\frac 12 Q(1, \bm v)$ and we set
$n_{\bar a}=\frac 12 Q(v, -\bm v)$,
$q_\perp =(0, \bm q_\perp)$ with $\bm q_\perp\cdot\bm v = 0$.
Given this basis, we calculate the initial evolution scale for
each particle as
\begin{equation}
  \label{eq:qtildeini}
  \tilde q_{\rm ini}^2 = (p_a+p_{\bar a})\cdot(p_a+n_{\bar a})
=\frac 12 Q^2 (1+v)\,. 
\end{equation}
We note that this is the most symmetric choice of initial conditions
(see \cite{NewVariables}).  In the $\qq$--case, this choice starts the
evolution of quark and antiquark at the same scale.  We could as well
choose another pair of evolution scales.  If we do so, however, we
make sure that the phase space region of soft gluon emission is
covered uniquely and smoothly with the radiation from the two partners
\cite{NewVariables}.  For later kinematic reconstruction we have to
store the momenta $p_i$ of the outgoing partons at this stage.

\subsection{Parton splittings and kinematics}
\label{sec:splittings}
Starting from the evolution scale $\tilde q_i = \tilde q_{\rm ini}$ we
now carry out the parton shower evolution for each final state
particle separately.  For every possible splitting $a\to bc$ of
particle $a$ we determine the scale of the next branching $\tilde q_{i+1}$
based on the Sudakov form factor
\begin{equation}
  \label{eq:suda}
  S_{ba}(\tilde q_{i}, \tilde q_{i+1}) 
  = \frac{\Delta_{ba}(\tilde q_c, \tilde q_i)}{
    \Delta_{ba}(\tilde q_c, \tilde q_{i+1})}\,, 
\end{equation}
where 
\begin{equation}
  \label{eq:delta}
  \Delta_{ba}(\tilde q_c, \tilde q) = 
  \exp\left\{-\int_{\tilde q_c}^{\tilde q} \!\frac{d\tilde q^2}{\tilde q^2}\,
  \int\! dz\, \frac{\alpha_S(z, \tilde q)}{2\pi}P_{ba}(z,\tilde q)
  \Theta({\bm p_\perp>0})\right\}\,.
\end{equation}
$\tilde q_c$ is the lower cutoff of the parton shower which, by
default, is taken to be the nonperturbative gluon mass $m_g=750\,$MeV.
$\alpha_S(z, \tilde q)$ is the running coupling in the case of QCD
evolution and generally depends on the evolution scale and momentum
fraction.  We choose $z(1-z) \tilde q$ as the argument of the running
coupling which reduces to the transverse momentum $\bm q_\perp$ in the
massless case.  $P_{ba}(z,\tilde q)$ are the quasi collinear splitting
functions that depend on the evolution scale in the case of massive
partons~\cite{HeavySplitting}.  For QCD branchings they are
\begin{align}
  P_{qq}(z, \tilde q) &= C_F\left[\frac{1+z^2}{1-z} 
  -\frac{2m_a^2}{z(1-z)\tilde q^2}\right]\,, \\
  P_{qg}(z, \tilde q) &= T_R\left[1-2z(1-z)
  +\frac{2m_a^2}{z(1-z)\tilde q^2}\right]\,,\\
  P_{gg}(z, \tilde q) &= C_A\left[\frac{z}{1-z} 
    + \frac{1-z}{z} + z(1-z)\right]\,.
\end{align}
Similarly, for the branching $q\to q\gamma$, ignoring the parton mass,
we have,
\begin{equation}
  \label{eq:qgammasplit}
  P_{qq}^\gamma(z, \tilde q) = e_a^2\frac{1+z^2}{1-z}\,,
\end{equation}
with $e_a$ being the electric charge of the parton in units of the
elementary charge.  Of course we have to take the fine structure
constant $\alpha_{\rm em}$ in eq.~(\ref{eq:delta}) in this case.
$\Theta({\bm q_\perp>0})$ limits the phase space to the region where
it is possible to reconstruct the transverse momentum $\bm p_\perp$
from the evolution variables $(\tilde q, z)$, which is a complicated
and implicit function in our case.  However, using the veto algorithm
described below we do not need to know the phase space boundary
explicitly.

The evolution variables $\tilde q$ and $z$ determine the kinematics of
the parton shower.  The momentum fraction $z$ is simply the ratio of
the Sudakov variables $\alpha$ in eq.~(\ref{eq:sudakovbasis}) for the
parent and daughter parton,
\begin{equation}
  \label{eq:zdef}
  \alpha_{i+1} = z \alpha_i\,.
\end{equation}
Based on its meaning in the quasi-collinear limit, $\tilde q$
determines the relative transverse momentum as 
\begin{align}
  \label{eq:ptdef}
  |\bm p_\perp| &= \sqrt{(1-z)^2 (\tilde q^2 -\mu^2) - zQ_g^2} 
  &\qquad (\text{quark branching}),\\
  \label{eq:ptdef2}
  |\bm p_\perp| &= \sqrt{z^2(1-z)^2 \tilde q^2 -\mu^2} 
  &\qquad (\text{gluon branching})\;,
\end{align}
where 
\begin{equation}
  \mu = \max(m_a, Q_g)\label{eq:mudef}\;,
\end{equation}
when a quark of mass $a$ is involved in the branching and simply
$\mu=Q_g$ for the splitting $g\to gg$. Here we have introduced the
cutoff $Q_g$ in order to regularize the soft gluon singularities in
the splitting function.  The relative transverse momentum $\bm
p_\perp$ is related to the Sudakov variables (\ref{eq:sudakovbasis})
of the parton branching as
\begin{equation}
  \label{eq:ptrelation}
  \bm p_\perp = \bm q_{\perp i+1} - z \bm q_{\perp i}\,.
\end{equation}

From eqs.~(\ref{eq:ptdef}) and (\ref{eq:ptdef2}) we immediately get
the phase space constraint for $\bm p_\perp$ in eq.~(\ref{eq:delta}).
We require $z$ to correspond to a real value of $\bm p_\perp$.
For gluon splittings we explicitly obtain 
\begin{equation}
  \label{eq:zlimits}
  z_-<z<z_+, \qquad z_\pm =
  \frac{1}{2}\left(1\pm\sqrt{1-\frac{4\mu}{\tilde q}}\right)\quad
  \text{and}\quad\tilde q>4\mu\;. 
\end{equation}
For quark splittings the phase space boundary is the solution
of a cubic equation but the allowed $z$ range always lies within
\begin{equation}
  \label{eq:zlimit2}
  \frac{\mu}{\tilde q} < z < 1-\frac{Q_g}{\tilde q}\,.
\end{equation}
Therefore it is simplest to generate $z$ within this range and reject
those values that lie outside phase space. 
Finally, it takes an azimuthal angle $\phi$, which is currently chosen
randomly and may later be related to spin correlations \cite{PetersSpin},
to complete the four-momenta of the parton shower in a final kinematic
reconstruction.

\subsection[Parametrization of $Q_g$]{Parametrization of \boldmath $Q_g$}
\label{sec:qgpara}

We introduced the cutoff parameter $Q_g$ quite naturally as a gluon
virtuality into the shower kinematics. Considering the phase space
that is available to the parton shower, we would expect a natural
threshold in $\tilde q$ of the order of $m_q+Q_g$.  In contrast, we
find from eq.~(\ref{eq:ptdef}) that the actual threshold behaves
approximately as $Q_{\rm thr}= 1.15(m_q+2Q_g)$.  Hence, we find that,
particularly for heavy quarks the phase space limit is well above our
expectation.

There is no reason why $Q_g$ should be kept as the same parameter for all
quark flavours.  Therefore, we have chosen to parametrize the
threshold for different flavours in terms of a unique parameter
$\delta$ as
\begin{equation}
  \label{eq:deltap}
  Q_g = \frac{\delta - 0.3\,m_q}{2.3}\;, 
\end{equation}
which leaves us with a threshold $Q_{\rm thr} = 0.85\,m_q+\delta$ for
all flavours\footnote{In principle, the coefficient of $m_q$
could be a model parameter.}. 
The resulting phase space in $\tilde q$ and $z$ is then
as shown in fig.~\ref{fig:qtzps}.  We show the dependence of our
results on the parameter $\delta$ in most of the plots of
sec.~\ref{sec:results}.  In the case of gluon splitting $m_q$ is the
mass of the splitting products, i.e.\ the quark mass in case of a
$g\to q\bar q$ splitting or $m_q=0$ in $g\to gg$ splitting.

\FIGURE{
  \epsfig{file=thr.10}
  \epsfig{file=thr.11}
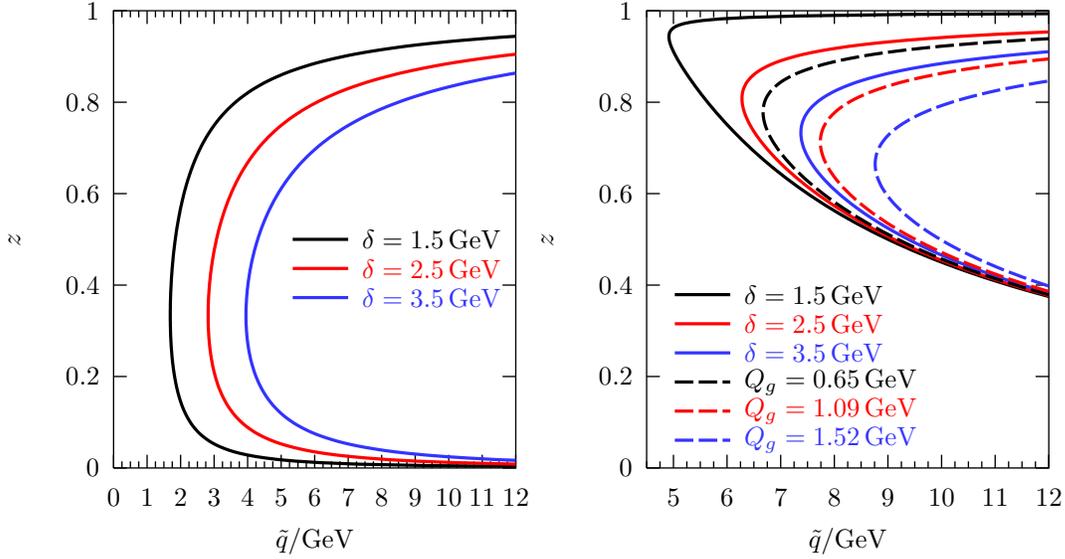
  \caption{Available phase space of light (left) and $b$--quarks 
    (right) for $q\to qg$ splitting for various values of $Q_g$ and
    depending on the parametrization in terms of $\delta$,  
    eq.~\eqref{eq:deltap}.  The dashed lines on the right correspond
    to the same $Q_g$ values as for the light quarks.}
  \label{fig:qtzps}
}

\subsection{Single branching process}
\label{sec:branching}
For timelike (i.e.\ final state) branchings, given an initial scale
$\tilde q_i$, the Sudakov form factor eq.~(\ref{eq:suda}) gives the
probability for no branching above the scale $\tilde q_{i+1}$.  Hence,
$1-S_{ab}(\tilde q_i, \tilde q_{i+1})$ is the probability for the next
branching to happen above $\tilde q_{i+1}$ and its derivative with
respect to $\tilde q_{i+1}$ is the probability density for the next
branching to happen at the scale $\tilde q_{i+1}$.

We sample the next branching scale with the veto algorithm. We
overestimate the integrands as follows.  We take the absolute maximum
of the relevant coupling $\alpha_{\rm max}$ as this is generally a
very slowly varying function.  The splitting functions are
overestimated by
\begin{align}  
  g_{qq}(z) &= \frac{2C_F}{1-z}\,,\\
  g_{qg}(z) &= T_R\,,\\
  g_{gg}(z) &= C_A\left[\frac{1}{1-z} + \frac{1}{z}\right]\,,\\
  g_{qq}^\gamma(z) &= \frac{2e_a^2}{1-z}\,,
\end{align}
in such a way that their integrals $G_{ba}(z)$ are invertible
functions.  Furthermore, they do not depend on $\tilde q$ anymore.
The phase space in $z$ is overestimated by taking the maximum value of
the evolution scale, $\tilde q_i$.  From this, we calculate the limits
in $z$ from eq.~(\ref{eq:zlimits}) or (\ref{eq:zlimit2}),
respectively.  As we can now easily integrate and invert the exponent
in eq.~(\ref{eq:delta}), we can sample values $\tilde q_s$ and $z_s$.
Then we subsequently apply vetoes with weights
\begin{equation}
  \label{eq:weights}
  w_1 = \Theta(\bm p_\perp>0)\,, 
  \quad w_2 = \frac{P_{ba}(z, \tilde q)}{g_{ba}(z)}\,,
  \quad w_3 = \frac{\alpha(z, \tilde q)}{\alpha_{\rm max}}\,.
\end{equation}
When all vetoes are passed, we have a scale $\tilde q_{i+1}= \tilde
q_s$ and a momentum fraction $z_s$ value. If not, we try to obtain a new
branching, now starting at scale $\tilde q_s$, repeating until we either
accepted a scale as the next branching scale or we obtain a scale
$\tilde q_s < \tilde q_c$ at which we cannot resolve a parton any further. 

In this way we calculate branching scales $\tilde q_{i+1}$ for every
possible splitting process for a given particle.  The splitting with
the largest scale of those above $\tilde q_c$ is then taken to be the
next branching.  In this way we can easily include any type of
branching.

\subsection{Angular ordering}
\label{sec:angord}
Once a parton is split its resolution scale $\tilde q_i$ is still
above the smallest resolution scale~$\tilde q_c$.  In order to have
angular ordering we now calculate the subsequent branchings of its
daughters as $\tilde q_{i+1}$ and $\tilde k_{i+1}$ with the conditions
\begin{equation}
  \label{eq:angord}
  \tilde q_{i+1} < z\tilde q_i
  \qquad \tilde k_{i+1} < (1-z) \tilde q_i\,.
\end{equation}
This branching process is repeated until no more daughter particles
are resolved at scales above the resolution scale $\tilde q_c$.  Note
that, for our choice of evolution variables, the parton shower is
terminated because there is no more phase space available at low
scales.  The lower limit of evolution is normally given by the soft
gluon cutoff $Q_g$ (or $\delta$) or the masses $m$ of the branching
particles (cf.\ fig.~\ref{fig:qtzps}).  However, when we choose very
small cutoffs $Q_g$, which are in principle allowed, we apply the
additional constraint $\tilde q>\tilde q_c$ on the shower termination.

\subsection{Soft matrix element correction}
\label{sec:softmec}
As explained in sec.~\ref{sec:hardmec} we explicitly populate the
`dead region' of the $\qqg$ phase space according to the correct QCD
matrix element.  We also improve parton shower emissions within the
shower regions of the phase space, as the parton shower might generate
relatively hard gluon emissions which are not within the domain of
validity of the quasi-collinear approximation anymore. 

In order to do so, we keep track of the relative transverse momentum
$\bm p_\perp$ (cf.\ eqs.~\eqref{eq:ptdef}, \eqref{eq:ptdef2}) that was
generated during the parton shower evolution of one jet.  Whenever we
find that this transverse momentum is the largest that was generated
during the evolution so far we apply a so-called soft matrix element
correction~\cite{Mike}.

We consider all previous gluon emissions as being infinitely soft in
comparison to the one we are testing.  This allows us to compute the
three-body (i.e.\ $\qqg$) variables $(x, \bar x)$ from the parton
shower variables $(\tilde q, z)$ and the respective Jacobian.  Then we
compare a random number with the ratio of the true matrix element to
the parton shower approximation and reject the branching if the ratio
is smaller.  Clearly this requires the parton shower emission
probability to be larger that the matrix element everywhere in phase
space, which is true for this process and our choice of evolution
variables.

\subsection{Reconstruction of kinematics}
\label{sec:kinreco}
As we generate the parton splittings $i\to (i+1)k$ we can calculate
the $\alpha_{i+1}$--component and the transverse momentum of the
daughter parton $(i+1)$ using eqs.~(\ref{eq:zdef}) and
(\ref{eq:ptrelation}).  The respective variables of the second
particle $k$ are simply obtained using $1-z$ and the difference of the
transverse momenta of $i$ and $(i+1)$.  However, we can only
reconstruct the $\beta$ variables when we know the virtuality of
each particle.  This is done recursively once the parton shower evolution
has terminated.  The final-state particles are put on their
constituent mass shells and then we obtain the beta components from
\begin{equation}
  \label{eq:beta}
  \beta = \frac{q^2 + \bm q_\perp^2 -\alpha^2 p^2}{2\alpha p\cdot n}\;,
\end{equation}
where $p$ and $n$ are the Sudakov basis vectors of the shower. These
were determined in the initial phase and remain fixed for each jet.

After the completion of the shower evolution of every parton $j$ involved in
the hard process, the jet parent partons are not on their mass shells
$p_j^2=m_j^2$ anymore.  Instead, they have acquired  virtualities
$q_j^2$.  If the original momenta were given as $p_j = (\sqrt{\bm
  p_j^2+ m_j^2}, \bm p_j)$ in the centre-of-mass frame of the
hard process, we want to preserve the total energy in this frame,
\begin{equation}
  \label{eq:roots}
  \sqrt{s}= \sum_{j=1}^n \sqrt{m_j^2 + \bm p_j^2}\;,
\end{equation}
while we want to keep the sum of spatial momenta vanishing.  As the
jet parents have momenta $q_j = (E_j, \bm q_j)$ after the showering,
we need some way to restore momentum conservation in a way that most
smoothly preserves the internal properties of each jet.

The simplest way to do this so-called ``momentum reshuffling''
is to rescale the momentum of each jet with
a common factor $k$ that is determined implicitly from
\begin{equation}
  \label{eq:kdetermine}
  \sqrt{s}= \sum_{j=1}^n \sqrt{q_j^2 + k\bm p_j^2}\,.
\end{equation}
Then, for every jet we determine a Lorentz transformation such that
\begin{equation}
  \label{eq:boost}
  q_j = (E_j, \bm q_j) \stackrel{\text{boost}}{\longrightarrow} 
  q_j' = (E_j', k\bm p_j)\,. 
\end{equation}
Typically the rescaling parameters $k$ are very close to unity and
hence the resulting angles and boost parameters in
eq.~(\ref{eq:boost}) are small.

\section{The cluster hadronization in detail}\label{sec:hadro}
The cluster hadronization has two main steps. The first is the cluster
formation, in which all of the colour connected partons created in the
shower are combined to form clusters which are colour singlets. The
other step is the decay of these colour singlet clusters into hadrons.
The new model presented here only changes the second stage, how the
cluster decays.  The process of cluster formation remains identical to
$\HW$, with the same set of parameters.

\subsection{Cluster formation}
The gluons in the partonic final state are split non-perturbatively into $\qq$ 
pairs. The choice of flavour is between the $u,d$ and $s$ flavours. The 
splitting is done with a simple isotropic decay where the gluon is given an
effective gluon mass, $m_g>2m_q$. The default value for $m_g$ is $0.75\,$GeV.

Once we have a state of all on-shell quarks, the colour partners are combined
into clusters. Owing to the colour-preconfinement property of the parton
shower~\cite{preconf}, the cluster mass distribution is independent of
the nature and energy of the hard process to a good approximation.
This can be seen in Figure \ref{fig:clusterDist} for light ($uds$)
quark clusters and clusters containing a $b$ quark separately.   

\FIGURE{ 
  \epsfig{file=cluana3.8,scale=0.9}\hfill
  \epsfig{file=cluana3.6,scale=0.9}
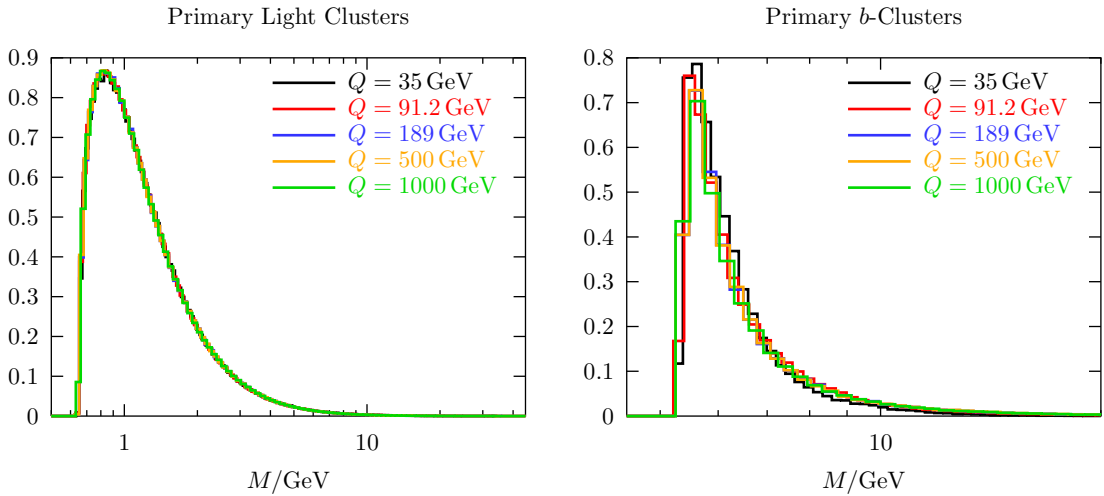
  \caption{Primary cluster mass distribution in $\ee$ annihilation at
    various centre-of-mass energies $Q$ for clusters containing only light
    quarks (left) and a $b$ quark (right).}
  \label{fig:clusterDist}
}

The hadronization model in $\HW$ and $\hpp$ also has a stage where
some of these clusters are decayed into two new clusters, rather than
directly to hadrons. This step is called cluster fission. The mass of
a cluster is given by $M^2 = p^2$, where $p$ is the momentum of the
cluster. The cluster $C$ is decayed into two new clusters $C_1, C_2$
if this mass does not satisfy the condition
\begin{equation}
M^\Clp < \Clm^\Clp + \Sigma_c^\Clp,
\end{equation}
where $\Clp$ and $\Clm$ are parameters of the model and $\Sigma_c$ is
the sum of the masses of the constituent partons which form the
cluster. If a cluster does decay into two new clusters, a flavour is
drawn from the vacuum. Again this is drawn from the $u,d$ and $s$
flavours.  The mass of cluster $i$ is drawn from the following
distribution
\begin{equation}\label{eq:Mdist}
M_i = \left[ \left( M^P - (m_i+m_3)^P\right)r_i + (m_i+m_3)^P \right]^{1/P},
\end{equation}
where $m_i$ is the mass of the constituent quark from the original
cluster that is going into the new cluster and $m_3$ is the
constituent mass of the flavour that was drawn from the vacuum. Here
$P$ is a parameter of the model and $r_i$ is a random number.
The value of $P$ is
given by $\texttt{PSPLT(1)}$ if parton $i$ is of $u,d,s$ or $c$
flavour and by $\texttt{PSPLT(2)}$ for a cluster where parton $i$ is
of $b$ flavour.  The two masses are also correlated by the constraint
that $M_1+M_2 \leq M$.  If this constraint is violated, a new flavour
is drawn from the vacuum and two new cluster masses are drawn from the
distribution (\ref{eq:Mdist}).  The decay kinematics is determined in
the rest frame of $C$, as the original constituent quarks continue
their movement in the same direction also when they are boosted into
the rest frame of the new clusters $C_1, C_2$.  As all masses are
given, the momenta of clusters and constituents are determined.

\subsection{Cluster decays}
The last stage of the hadronization is the cluster decays. The problem
with the original \HW\ cluster decay model~\cite{Webber:1983if} can be
shown as follows.  The probability of accepting a decay of a cluster
with flavours $i,j$ into hadrons of type $a,b$ is
\begin{equation}
P(a_{i,q},b_{q,j}|i,j) = P_q P(a|i,q) P(b|q,j) P_{\rm PS}(a,b)\;.
\end{equation}
Here $P_q$ is the probability of drawing flavour $q$ from the vacuum and
$P_{\rm PS}$ is the probability due to phase space. The probabilities of 
interest are the other two. These have the form
\begin{equation}
P(a|i,j) = \frac{w_a}{N_{ij} M_{ij}},
\end{equation}
where $w_a$ is a hadron specific weight, $M_{ij}$ is the maximum weight of all 
the hadrons of flavour $i,j$ and $N_{ij}$ is the number of hadrons of flavour
$i,j$ in the model. We can see that the probabilities have a dependence on
$N_{ij}$.
As described in~\cite{kupco} this causes adverse side effects when new hadrons
are added to the model. If we consider adding a new hadron of flavour 
$u \overline{d}$, for example, it will generally be heavier than those
already present.  This will suppress the probability 
of choosing a lighter $u \overline{d}$ meson as $N_{u\bar d}$ is increased.
Therefore properties such as the charged to neutral pion ratio are controlled
heavily by how many hadrons of a particular flavour are in the model. 

To solve this problem a new construction of the probability was
created in~\cite{kupco}.  Instead of independently choosing the
flavour from the vacuum and then choosing the hadrons, this is all
combined into one distribution. The weight of one choice is
\begin{equation}
W(a_{i,q},b_{q,j}|i,j) = P_q w_a w_b P_{\rm PS}(a,b)\;.
\end{equation}
This gives the probability
\begin{equation}
P(a_{i,q},b_{q,j}|i,j) = \frac{W(a_{i,q},b_{q,j}|i,j)}{\sum_{q',c,d} 
  W(c_{i,q'},d_{q',j}|i,j)}.
\label{eq:kupco_prob}
\end{equation}

Because $P_{\rm PS}$ is zero for heavy decay modes only accessible modes 
influence the probabilities. Unfortunately, this solution has a new problem in 
that the ratio of mesons to baryons is dictated by the number of available 
particles. Since there are many more mesons then baryons the denominator in 
eq.~(\ref{eq:kupco_prob}) is quite large and the total probability of
choosing a 
baryonic decay mode is very small. So though this new approach is able to make 
quantities such as pion ratios independent of the number of hadrons in the
model it fails to produce the correct amount of baryons.

The solution implemented in \hpp\ is to treat the baryon sector
independently from the meson sector. This is done by re-interpreting
the parameter for the diquark weight, ${\texttt{Pwt}}_{\rm di}$, to be
the parameter that controls the frequency of drawing independently
from the baryon sector.  This is expressed as
\begin{equation}
P_{\rm B} = \frac{{\texttt{Pwt}}_{\rm di}}{{\texttt{Pwt}}_{\rm di}+1}.
\end{equation}
So there is a probability $P_B$ of choosing only from the baryon sector and a
probability $P_M = 1-P_B$ of choosing from the meson sector. The actual choice 
of hadrons is then made according to the probability
\begin{equation}
P(a_{i,q},b_{q,j}|i,j) = \frac{W(a_{i,q},b_{q,j}|i,j)}{\sum_{M/B} 
  W(c_{i,q'},d_{q',j}|i,j)}.
\end{equation}
where the sum over $M/B$ indicates only summing over the flavours that produce 
either mesons or baryons.

\subsection{Hadron decays}
Most of the hadrons created in the cluster hadronization are not stable and 
need to be decayed. At present, the decays in \hpp\ are done in the same way
as in \HW. Most decays are treated as simple $n$-body isotropic decays.
Weak decays are done by either free particle $V-A$ matrix elements or
bound quark $V-A$ matrix elements. Examples of these are
 $\tau^- \to e^- \overline{\nu}_e \nu_{\tau}$ and
$K^- \to e^- \overline{\nu}_{e} \pi^0$, respectively.

Heavy hadrons, such as B mesons, are sometime decayed into partonic states. 
These states are then fed back into the shower and are re-showered and 
re-hadronized. There are two different types of heavy partonic decays.
One is a weak decay, for which we use the same free or bound $V-A$
matrix elements as for the light mesons. This would occur, for example, in the 
decay $B^0  \rightarrow \overline{s} c \overline{c} d$. There are also
quarkonium decays that have gluons as decay products, for example
the decay $\eta_c \rightarrow gg$. These decays are done using the
appropriate positronium matrix elements.

\section{Observables}

\subsection{Observables considered}
\label{sec:list}
We considered the following observables in our study. 
\label{sec:obs}
\paragraph{Event shape variables:} Event shape distributions have
been measured to very high accuracy at LEP and aim at resolving the
properties of the parton shower quite deeply.  In particular we have
chosen the thrust ($T$), thrust major ($M$), thrust minor ($m$) and
oblateness ($O$) as they are the most commonly used.  In addition we
look at the $C$--parameter and $D$--parameter as they are more
sensitive to multijet events.  We also look at the sphericity and
planarity ($S, P$), which are calculated from the quadratic momentum
tensor and therefore put more emphasis on 2-jet like events.
Furthermore we look at the wide and narrow jet broadening measures
($B_{\rm max}$, $B_{\rm min}$), which are more sensitive to the
transverse jet structure.

\paragraph{Jet multiplicity:} The multiplicity of (mini-)jets in
$\ee$-collisions for different values of the jet resolution $\ycut$.
We use the Durham-- or $k_\perp$--clustering scheme \cite{Durham}
throughout the paper for jet observables. To be specific, for a given
final state the jet measure 
\begin{equation}
\label{eq:yij}
y_{ij} = \frac{2\min(E_i^2, E_j^2)}{Q^2}(1-\cos\theta_{ij})
\end{equation}
is calculated for every particle pair $(i, j)$. The particles with
minimal jet measure are clustered such that the momentum of the
clustered pseudo-particle is the sum of the four-momenta of the
constituents. The jet multiplicity is then the number of
pseudo-particles remaining when all $y_{ij}>\ycut$.  This inclusive
observable has been predicted and measured at LEP energies and will
test the dynamics of the parton shower as well as the interface
between parton shower and hadronization. We use the
\texttt{KtJet}-package \cite{ktjet} that implements the above
jet-finding algorithm in C++ and have written a simple wrapper around
it in order to use it with our own particle record.

\paragraph{\boldmath Jet fractions and $Y_n$:} A closer look
`into' the jets is provided by considering the rates of jets at a
given value of $\ycut$ in the Durham scheme, $R_n=N_{n-{\rm
    jet}}/N_{\rm evts}$ for $n=2$ up to $n=6$ jets.  We also look at
the distributions of $Y_n$, the $\ycut$-values at which an $n+1$ jet
event is merged into an $n$-jet event in the Durham clustering scheme.
Here we look at $n=2$ up to $n=5$.  These distributions will not only
probe the dynamics of the parton shower but also the hadronization
model: at the lowest values of $\ycut \sim (\tilde q_c/Q)^2$ the
dynamics is dominated by the latter.

\paragraph{Four--jet angles:} A very interesting set of observables are
the distributions of the angles between jets in four--jet events,
$\chi_{\rm BZ}$, $\Phi_{\rm KSW}$, $\theta^*_{NR}$ and $\alpha_{34}$,
defined for example in ref.~\cite{Heister:2002tq}.  These angles are
expected to be sensitive to the accuracy of the simulation of
higher--order matrix elements.

\paragraph{Single particle distributions:}
Another interesting set of observables are the momentum distributions
of final-state particles with respect to the event orientation.  $y^T$
is the rapidity distribution with respect to the thrust axis,
$p_{\perp, {\rm in}}^T$ and $p_{\perp, {\rm out}}^T$ are the
respective transverse momenta within and out of the event plane,
defined by the thrust and thrust major axes.  Without any reference to
the event orientation, we look at the distribution of the momentum
fraction $x_p=2|\bm p|/Q$ and at $\xi_p = -\log x_p$ which displays
better the effects of soft gluon coherence at small $x_p$.
  
\paragraph{Identified particle spectra:}
We consider the
exclusive momentum distributions of $\pi^\pm$, $K^\pm$, $p, \bar p$
and $\Lambda, \bar\Lambda$.  These are generally expected to be
sensitive to the hadronization model.  In all cases except $\Lambda,
\bar\Lambda$ we can compare with data on the momentum distributions
from $uds$, $c$ and $b$ events separately.

\paragraph{Hadron multiplicities:} The
charged particle multiplicity distribution and the
average multiplicities of a wide range of hadron species
were taken to test the overall flow of quantum numbers through the
different stages of simulation.  The improved cluster hadronization
model can be tested thoroughly against these observables. 

\paragraph{B fragmentation function:} The energy fraction of
$B$-hadrons is taken as a test for the new parton shower which is
specifically designed to improve the description of heavy quark
observables with respect to the description in \HW.

\vspace{2mm}
The above list of observables has proven to be very useful to test
different domains of the available phase space of parameters and has
led us to important conclusions for the ongoing development of the
code for hadronic collisions.

\subsection{Analysis}
\label{sec:ana}

We have booked histograms for all the above distributions in the same
bins as the experimental data.  For a given bin $i$ we then compare
the data $D_i$ value with the \hpp{} Monte Carlo result $M_i$.  Given
the data errors $\delta D_i$ (statistical plus systematic) and Monte
Carlo errors $\delta M_i$ (statistical only), we can calculate a
$\chi^2$ for each observable. We keep the statistical error of the
Monte Carlo generally smaller than the experimental error.  In
distributions where the normalization is not fixed, such as momentum
spectra, we allow the normalization of the Monte Carlo to be free to
minimize $\chi^2$.  The normalization is then tested separately
against the average multiplicity.  In all other cases we normalize
histograms to unity.

As we do not want to put too much emphasis on a single observable or a
particular region in phase space where the data are very precise, in
computing $\chi^2$ we set the relative experimental error in each bin
to $\max(\delta D_i/D_i,5\%)$.  This takes into account the fact that
the Monte Carlo is only an approximation to QCD and agreement with the
data within 5\% would be entirely satisfactory.  The general trend for
the preferred range of a single parameter was however never altered by
this procedure.

After normalization the ratio 
\begin{equation}
  \label{eq:binratio}
  R_i = \frac{M_i-D_i}{D_i} \pm \left( \frac{\delta M_i}{D_i} 
    \oplus \frac{M_i \delta D_i}{D_i^2}\right)
\end{equation}
is computed for each bin in order to see precisely where the model
fails.  This ratio as well as the relative experimental error (yellow)
and the relative contribution of each bin to the $\chi^2$ of an
observable is plotted below each histogram.

\subsection{Strategy}
\label{sec:multcomment}
We have taken $\chi^2$ values for hadron multiplicities into account
in the same way as we weighted the event shapes.  In general the
multiplicities of individual particle species are sensitive to a
completely different set of parameters.  The general strategy was to
start from an initial set of hadronization parameters taken from \HW,
and to aim for a good value for the total number of charged particles
with reasonable values for the parton shower cutoff parameter
$\delta$ and the maximum cluster mass parameter $\texttt{CLMax}$.
Once these were fixed, the hadronization parameters that determine the
multiplicities of individual particle species were determined.
Following this we compared this `preferred' set of parameters with the
`initial' set from \HW{}.  The resulting parameter set is shown in
table \ref{tab:Herwig_defaults}. 

\TABLE[p]{
\begin{tabular}{lll}
\hline
Parameter & Preferred & Initial\\
\hline \hline
$\alpha_s(M_Z)$  & 0.118 & 0.114 \\
$\delta/$GeV     & 2.3 & ---\\
$m_g/$GeV        & 0.750 &---\\
$Q_{\rm min}$/GeV in $\alpha_s(Q_{\rm min})$\quad{} 
& 0.631\qquad{}
&---\\
\hline
$\Clm/$GeV                               & 3.2 & 3.35\\
$\Clp$                               & 2.0 &---\\
PSplt1 & 1&---\\
PSplt2 & 0.33 &---\\
B1Lim  & 0.0&---\\
ClDir1 & 1&---\\
ClDir2 & 1&---\\
ClSmr1 & 0.40 &---\\
ClSmr2 & 0.0&---\\
${\rm Pwt}_d$ & 1.0&---\\
${\rm Pwt}_u$ & 1.0&---\\
${\rm Pwt}_s$ & 0.85 & 1.0\\
${\rm Pwt}_c$ & 1.0&---\\
${\rm Pwt}_b$ & 1.0&---\\
${\rm Pwt}_{di}$ & 0.55 & 1.0\\
Singlet Weight & 1.0&---\\
Decuplet Weight & 0.7 & 1.0\\
\hline \hline
\end{tabular}
\caption{The preferred parameters for \hpp{}. The first group are
shower parameters, the second are all of the hadronization
parameters. In the third column we show initial values of our study,
taken from \HW{}, where these differ from the preferred values.}
\label{tab:Herwig_defaults}
}

\section{Results}
\label{sec:results}
We have chosen a wide range of observables in order to test different
aspects of the model.  Event shape variables and multiplicities are
considered in order to test the dynamical aspects of parton shower and
hadronization models, which are closely linked at their interface via
the parton shower cutoff parameter $\delta$.  Ideally, the two models
should merge smoothly at scales where $Q_g\sim 1\,$GeV.  All figures
shown at the end of the paper contain three sets of plots:
\begin{itemize}
\item the actual distribution.  The \hpp{} result is plotted as a
  histogram together with the experimental data points (top);
\item the ratio $R_i$ \eqref{eq:binratio} together with an error band,
  showing the relative experimental statistical and systematic
  errors (middle);
\item the relative contribution of each data point to the total
  $\chi^2$ of each plot (bottom).
\end{itemize}

\subsection{Hadron multiplicities}
Table \ref{tab:mult} shows the results of the new algorithm in
comparison with the old algorithm. The column labelled `Old Model' is
the result of using the old algorithm with the new shower variables.
The column \hpp{} is using the new algorithm with the new shower and
the last column, labeled Fortran, is using the Fortran \HW\ program
(version 6.5).  Data are combined and updated from a variety of
sources, see ref.~\cite{Knowles:1997dk}.  We see that, even before
systematic tuning, the overall results are better than those of \HW,
with fewer prediction that differ from the data by more
than three standard deviations (starred).

\TABLE{
\small
\centering
\begin{tabular}{llllll} \hline
Particle & Experiment & Measured & Old Model & \hpp{} & Fortran
\\\hline
\hline
All Charged & M,A,D,L,O & 20.924 $\pm$ 0.117  & $20.22^*$   & $20.814$     & $20.532^*$\\
\hline                                                                    
$\gamma$  & A,O & 21.27 $ \pm$ 0.6 & $23.03$                & $22.67$    & $20.74$ \\ 
$\pi^0$   & A,D,L,O & 9.59 $ \pm$ 0.33 & $10.27$            & $10.08$    & $9.88$ \\ 
$\rho(770)^0$ & A,D & 1.295 $ \pm$ 0.125 & $1.235$          & $1.316$    & $1.07$ \\ 
$\pi^\pm$ & A,O & 17.04 $ \pm$ 0.25 & $16.30$               & $16.95$    & $16.74$ \\ 
$\rho(770)^\pm$ & O & 2.4 $ \pm$ 0.43 & $1.99$              & $2.14$     & $2.06$ \\ 
$\eta$ & A,L,O & 0.956 $ \pm$ 0.049 & $0.886$               & $0.893$   & $0.669^*$\\ 
$\omega(782)$ & A,L,O & 1.083 $ \pm$ 0.088& $0.859$         & $0.916$    & $1.044$ \\ 
$\eta'(958)$ & A,L,O & 0.152 $ \pm$ 0.03 & $0.13$           & $0.136$   & $0.106$ \\ 
\hline                                                                     
$K^0$ & S,A,D,L,O & 2.027 $ \pm$ 0.025 & $2.121^*$          & $2.062$    & $2.026$ \\ 
$K^*(892)^0$ & A,D,O & 0.761 $ \pm$ 0.032 & $0.667$         & $0.681$   & $0.583^*$ \\ 
$K^*(1430)^0$ & D,O & 0.106 $ \pm$ 0.06 & $0.065$           & $0.079$   & $0.072$ \\ 
$K^\pm$ & A,D,O & 2.319 $ \pm$ 0.079 & $2.335$              & $2.286$    & $2.250$ \\ 
$K^*(892)^\pm$ & A,D,O & 0.731 $ \pm$ 0.058 & $0.637$       & $0.657$   & $0.578$ \\ 
$\phi(1020)$ & A,D,O & 0.097 $ \pm$ 0.007 & $0.107$         & $0.114$   & $0.134^*$ \\ 
\hline                                                                     
$p$ & A,D,O & 0.991 $ \pm$ 0.054 & $0.981$                  & $0.947$   & $1.027$ \\ 
$\Delta^{++}$ & D,O & 0.088 $ \pm$ 0.034 & $0.185$          & $0.092$   & $0.209^*$ \\ 
$\Sigma^-$ & O & 0.083 $ \pm$ 0.011 & $0.063$               & $0.071$   & $0.071$ \\ 
$\Lambda$ & A,D,L,O & 0.373 $ \pm$ 0.008 & $0.325^*$        & $0.384$   & $0.347^*$\\ 
$\Sigma^0$ & A,D,O & 0.074 $ \pm$ 0.009 & $0.078$           & $0.091$   & $0.063$ \\ 
$\Sigma^+$ & O & 0.099 $ \pm$ 0.015 & $0.067$               & $0.077$   & $0.088$ \\ 
$\Sigma(1385)^\pm$ & A,D,O & 0.0471 $ \pm$ 0.0046 & $0.057$ & $0.0312^*$ & $0.061^*$\\
$\Xi^-$ & A,D,O & 0.0262 $ \pm$ 0.001 & $0.024$             & $0.0286$   & $0.029$ \\ 
$\Xi(1530)^0$ & A,D,O & 0.0058 $ \pm$ 0.001 & $0.026^*$     & $0.0288^*$ & $0.009^*$\\ 
$\Omega^-$ & A,D,O & 0.00125 $ \pm$ 0.00024 & $0.001$       & $0.00144$   & $0.0009$ \\ 
\hline                                                                     
$f_2(1270)$ & D,L,O & 0.168 $ \pm$ 0.021 & $0.113$          & $0.150$   & $0.173$ \\ 
$f_2'(1525)$ & D & 0.02 $ \pm$ 0.008 & $0.003$              & $0.012$   & $0.012$ \\ 
$D^\pm$ & A,D,O & 0.184 $ \pm$ 0.018 & $0.322^*$            & $0.319^*$ & $0.283^*$ \\ 
$D^*(2010)^\pm$ & A,D,O & 0.182 $ \pm$ 0.009 & $0.168$      & $0.180$   & $0.151^*$\\ 
$D^0$ & A,D,O & 0.473 $ \pm$ 0.026 & $0.625^*$              & $0.570^*$ & $0.501$ \\ 
$D_s^\pm$ & A,O & 0.129 $ \pm$ 0.013 & $0.218^*$            & $0.195^*$ & $0.127$ \\ 
$D_s^{*\pm}$ & O & 0.096 $ \pm$ 0.046 & $0.082$             & $0.066$   & $0.043$ \\ 
$J/\Psi$ & A,D,L,O & 0.00544 $ \pm$ 0.00029 & $0.006$       & $0.00361^*$ & $0.002^*$\\ 
$\Lambda_c^+$ & D,O & 0.077 $ \pm$ 0.016 & $0.006^*$        & $0.023^*$ & $0.001^*$\\ 
$\Psi'(3685)$ & D,L,O & 0.00229 $ \pm$ 0.00041 & $0.001^*$  & $0.00178$   & $0.0008^*$\\ 
\hline
\hline
\end{tabular}
\caption{Multiplicities per event at 91.2 GeV. We show results 
  from \hpp{} with the implementation of the old cluster hadronization
  model (Old Model) and the new model (\hpp{}), and from \HW\ 6.5
  shower and hadronization (Fortran).  Parameter values used are given
  in table~\ref{tab:Herwig_defaults}.  Experiments
  are Aleph(A), Delphi(D), L3(L), Opal(O), Mk2(M) and SLD(S). The $*$
  indicates a prediction that differs from the measured value by
  more than three standard deviations.
  \label{tab:mult}} }

We also considered the distribution of the charged particle
multiplicity in comparison to OPAL data~\cite{Acton:1991aa} and find
fairly good agreement (fig.~\ref{fig:nchdist}), although with some
excess at low multiplicity.

\subsection{Jet multiplicity}
In fig.~\ref{fig:njetqg} we show the average number of jets $\langle
n_{jets}\rangle$ at the $Z^0$--pole, as a function of the Durham jet
resolution $\ycut$, for various values of the cutoff parameter
$\delta$.  At the parton level (top left) the jet multiplicity varies
a lot as we go to small values of $\ycut$, saturating at the number of
partons that are present in a single event.  The order of magnitude of
the visible saturation scales is characterized for each flavour by the
different cutoff values $Q_g$ as $y_{\rm sat} = Q_g^2/Q^2$ (cf.\ 
\eqref{eq:deltap}).  During hadronization, low parton multiplicities
lead to large mass clusters which tend to decay into low mass clusters
below a given cutoff mass, which is fixed to its default value
throughout the current section.  Fig.~\ref{fig:njetqg} (top right)
shows that the hadronization compensates for lower partonic
multiplicities, giving a result insensitive to $\delta$ at the hadron
level.  In other words, we have a smooth interface between
perturbative and non-perturbative dynamics between the lower end of
the parton shower on one side and the cluster hadronization model on
the other side.  On the hadron level we describe LEP data from OPAL
\cite{OPALnjet} well.

In order to test the sensitivity of our model against the variation of
the c.m.\ energy, we calculate the jet multiplicities at PETRA and LEP II
energies as well (fig.~\ref{fig:njetqg}, bottom). The comparison to
JADE \cite{JADEnjet} and OPAL \cite{OPALnjet} data shows a good
agreement.  In all runs we applied the same cutoffs on the energy of
the partonic subsystem as did the experiments.

The additional curves in fig.~\ref{fig:njetqg} show predictions for
the jet multiplicity \cite{NjetPrediction} from the resummation of
leading logarithms.  Note that the parameter $\Lambda_{\rm QCD}$
in the resummed calculation is not $\Lambda_{\overline{\rm MS}}$.
We see that for the value $\Lambda_{\rm QCD}=500$ MeV there is
good agreement with the data and the \hpp\ result throughout
the perturbative region, $\ycut > 10^{-4}$.

\FIGURE[p]{
  \epsfig{file=es.79,scale=1}
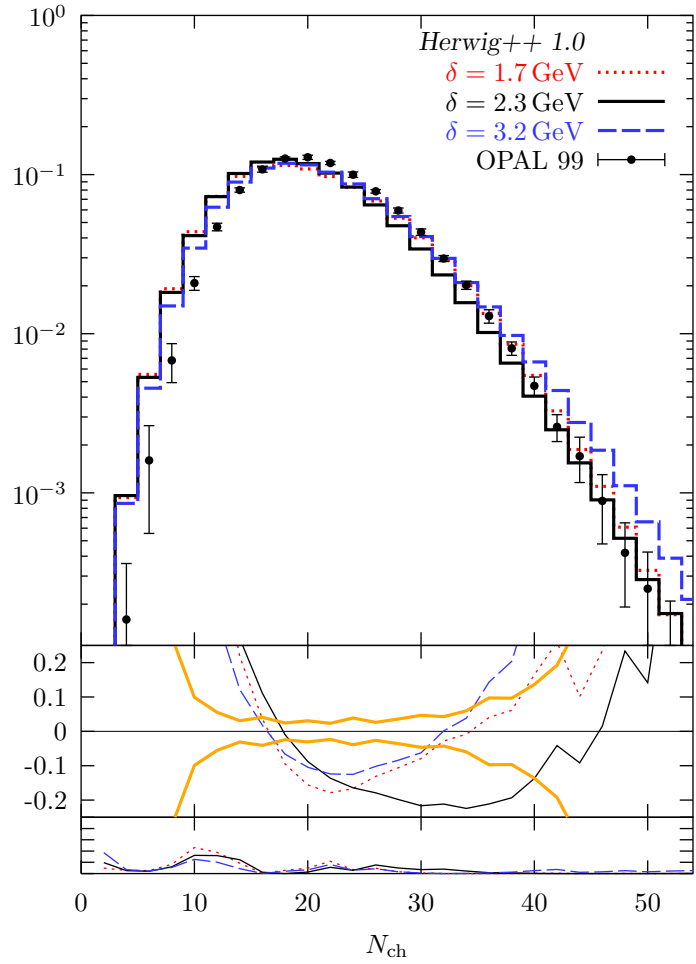
  \caption{The distribution of the charged particle multiplicity.
 The three panels here and in figs.~5--15 are explained in sec.~6.
    \label{fig:nchdist}}
}
\FIGURE[p]{
  \epsfig{file=njets2.4,scale=0.8}
  \epsfig{file=njets2.5,scale=0.8}
  \epsfig{file=njets2.7,scale=0.8}
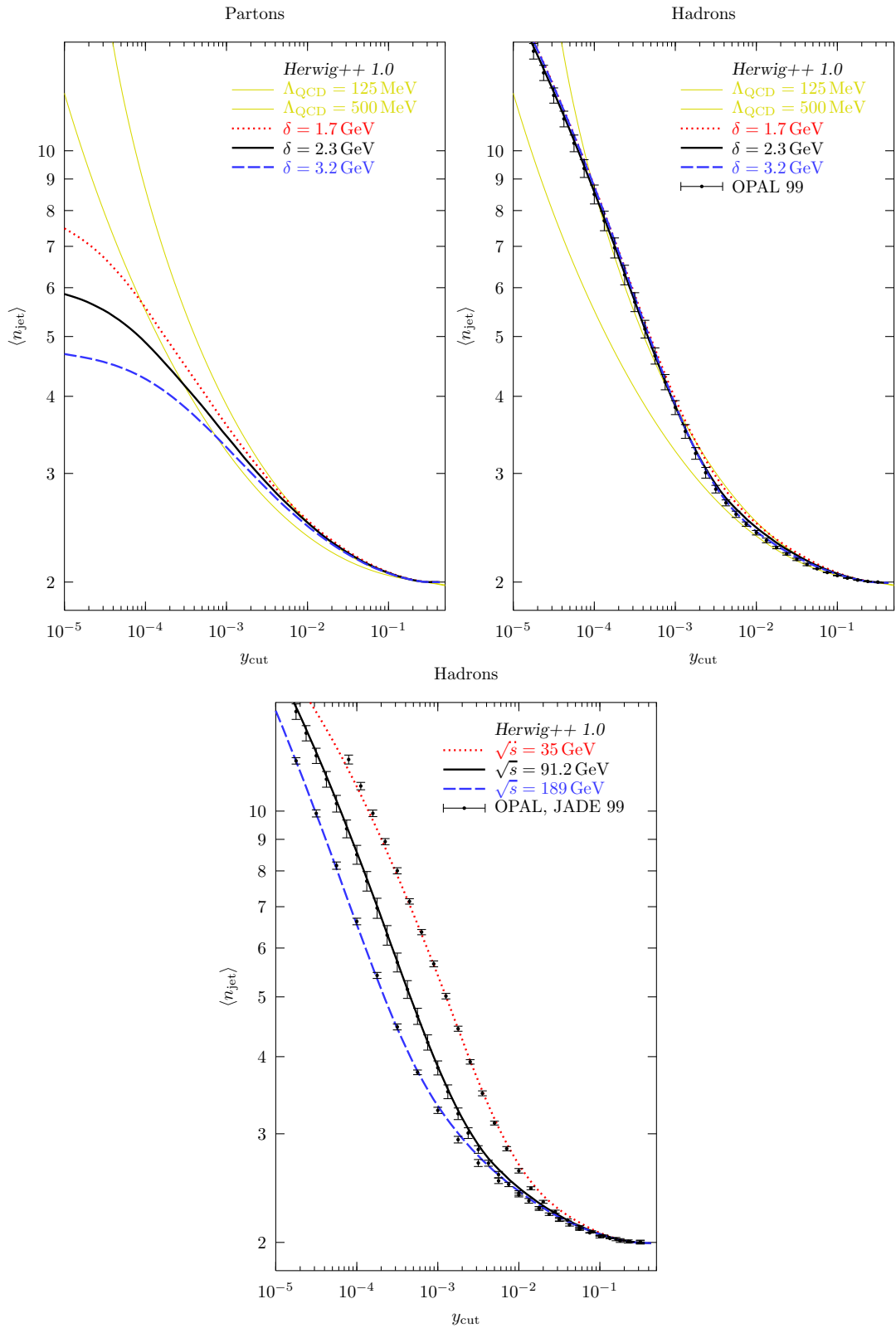
  \caption{Jet multiplicities for different values of the cutoff
    parameter $\delta$ and different c.m.\ energies.\label{fig:njetqg}}
}

\subsection[Jet rates and $Y_n$ distributions]{Jet rates and \boldmath 
  $Y_n$ distributions}

Another set of observables that is known to be well-described at LEP
energies are the fractions of $n$-jet events at a given $\ycut$ in the
Durham scheme.  In fig.~\ref{fig:ratesqg} we compare the results from
\hpp{}{} with LEP data from \cite{OPALnjet} and find good agreement.
On the hadron level these predictions are not very sensitive to the
cutoff parameter $\delta$.  The 5-jet distribution is not shown and
$R_6$ is the rate of $\ge 6$-jet events.

The Durham $Y_n$ distributions in fig.~\ref{fig:Ynqg} are histograms
of those $\ycut$--values at which an $n+1$--jet is merged into an
$n$--jet event in the Durham jet clustering scheme. We may say that
they resolve more internal structure of the jets than the $n$--jet
rates.  Still, the agreement between model and data is quite good,
although there is a tendency (which was also present in \HW) to
exceed the data at low $Y_n$.

\FIGURE[p]{
  \epsfig{file=es.61,scale=0.8}
  \epsfig{file=es.63,scale=0.8}\\[5pt]
  \epsfig{file=es.65,scale=0.8}
  \epsfig{file=es.69,scale=0.8}
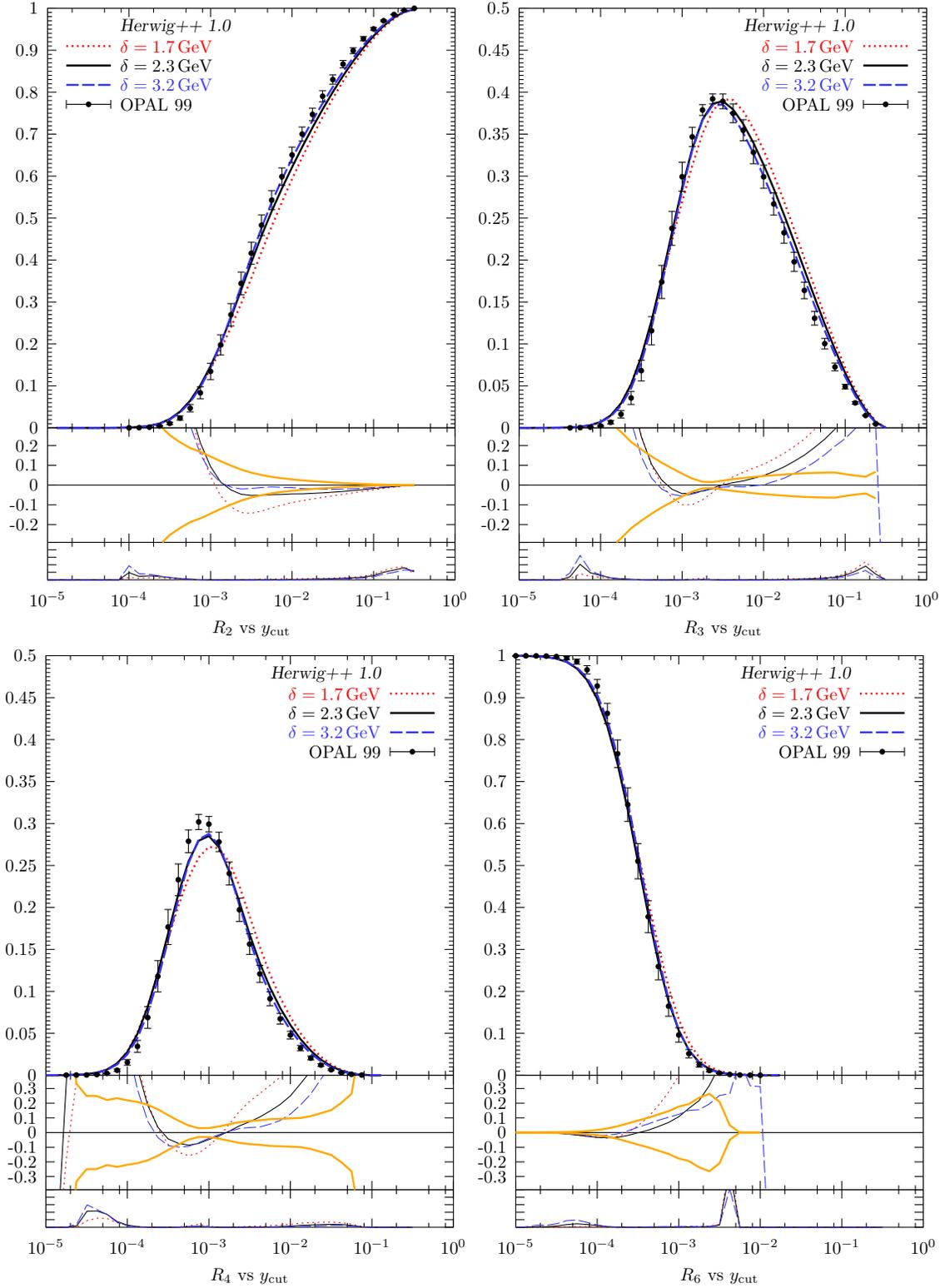
  \caption{Jet rates in the Durham algorithm for different values of
  the cutoff $\delta$.\label{fig:ratesqg}} }

\FIGURE[p]{
  \epsfig{file=es.51,scale=0.8}
  \epsfig{file=es.53,scale=0.8}\\[5pt]
  \epsfig{file=es.55,scale=0.8}
  \epsfig{file=es.57,scale=0.8}
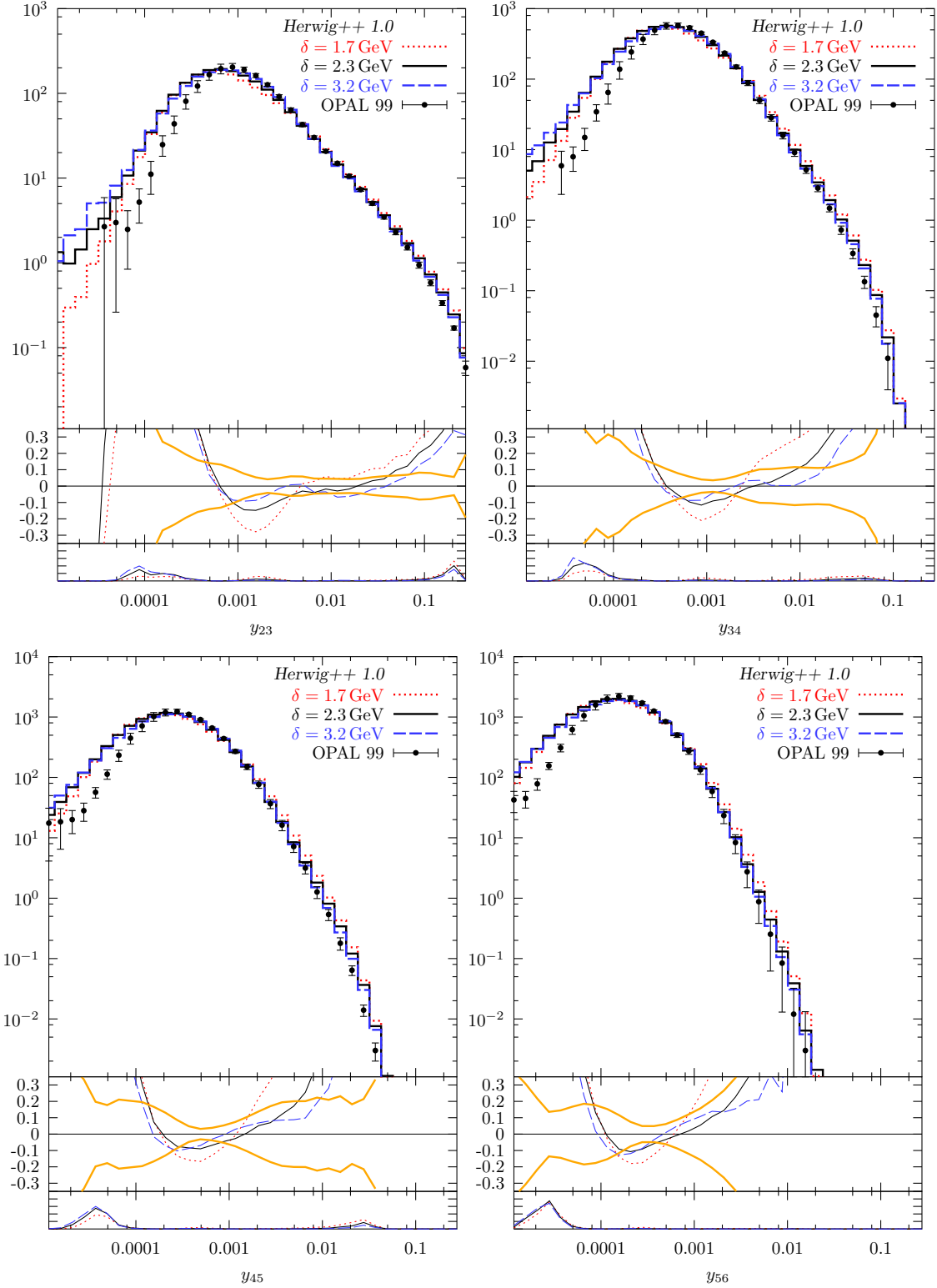
  \caption{Durham $Y_n$ distributions for different values of the 
    cutoff $\delta$.\label{fig:Ynqg}}
}

\subsection{Event shapes}
In order to test the dynamics of the parton shower in \hpp{} in more
detail we consider a set of commonly used event shape variables.  Not
only the collinear region of the parton shower is probed in greater
detail but also the regions of phase space which are vetoed as matrix
element corrections.  We compare all results to DELPHI data
\cite{DELPHIshapes}.

In fig.~\ref{fig:tocd} we show the distribution of thrust,
thrust-major and thrust-minor.  These variables are all obtained from
a linear momentum tensor.  The thrust distribution is shown with and
without matrix element corrections switched on. The prediction without
matrix element corrections is very much better than that of \HW, owing
to the improved shower algorithm.  It is interesting that the matrix
element corrections seem to generate almost too much transverse
structure, leading to event shapes that are less two-jet-like.  On the
other hand, there is also a slight excess of events close to the
two-jet limit.

It is remarkable how well distributions like $C$ and $D$ parameter
(fig.~\ref{fig:spcd}) which are sensitive to three- and four-jet-like
events are described by our model even though we are limited to three
jet matrix elements plus showers.  Here again we have in fact a small
excess at high values.

We show also in fig.~\ref{fig:spcd} the distributions of sphericity
and planarity, which are obtained from a quadratic momentum tensor
and therefore put more emphasis on high momenta.
As was the case for the thrust-related distributions, we tend to
have slightly wide events. In addition we consider the jet
broadening measures $B_{\rm max}$ and $B_{\rm min}$
and the hemisphere jet masses (fig.~\ref{fig:bmaxbmin}).
In all cases the agreement between model
and data is good.

\FIGURE[p]{
  \epsfig{file=es.0,scale=0.79}
  \epsfig{file=es.1,scale=0.79}\\[5pt]
  \epsfig{file=es.3,scale=0.8}
  \epsfig{file=es.5,scale=0.8}
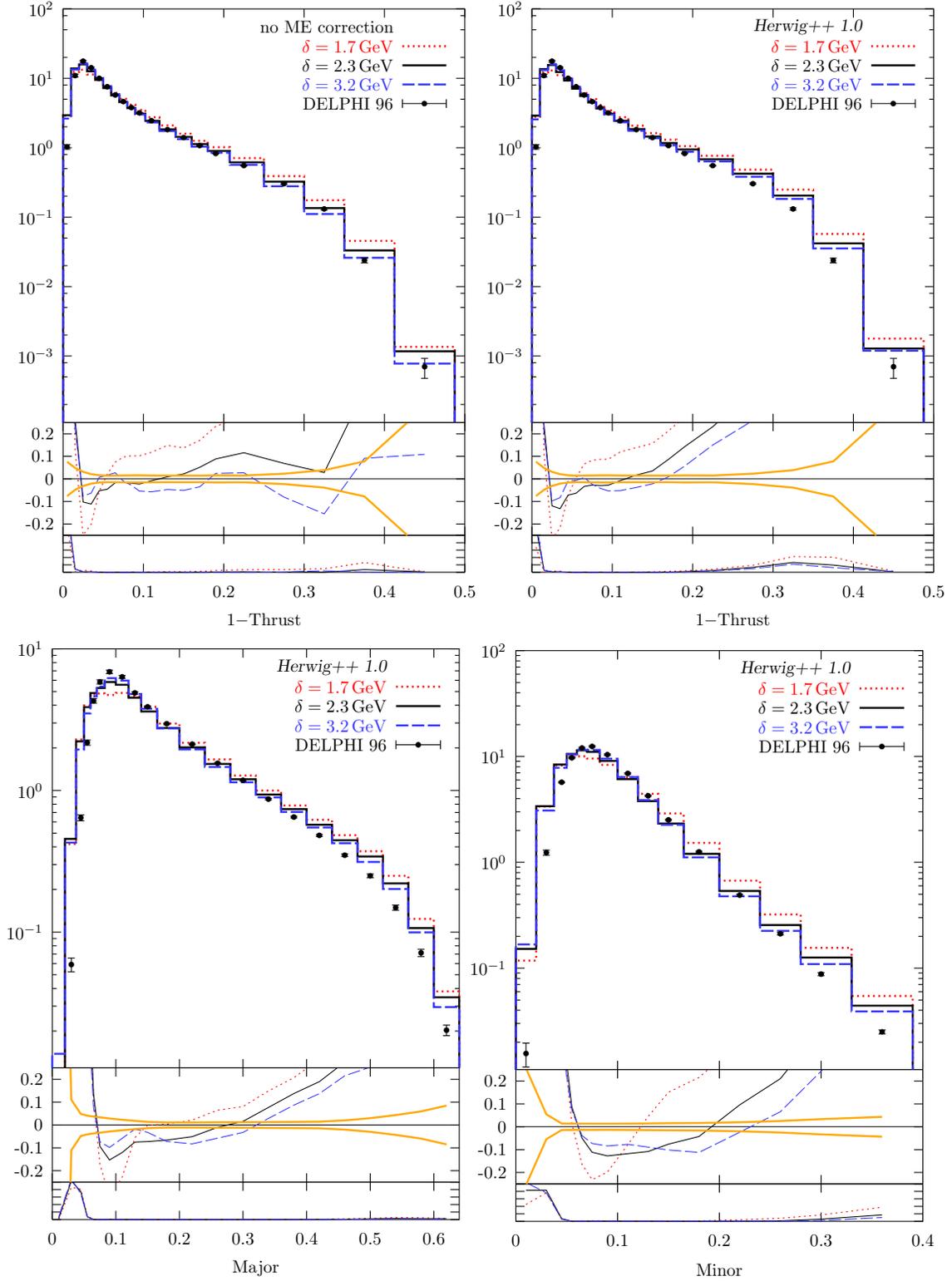
  \caption{Thrust without (top left) and with (top right) matrix
    element corrections switched on, thrust major and thrust minor
    (bottom).
    \label{fig:tocd}} }

\FIGURE[p]{
  \epsfig{file=es.9,scale=0.8}  
  \epsfig{file=es.11,scale=0.8}\\[5pt]
  \epsfig{file=es.15,scale=0.8}  
  \epsfig{file=es.17,scale=0.8}  
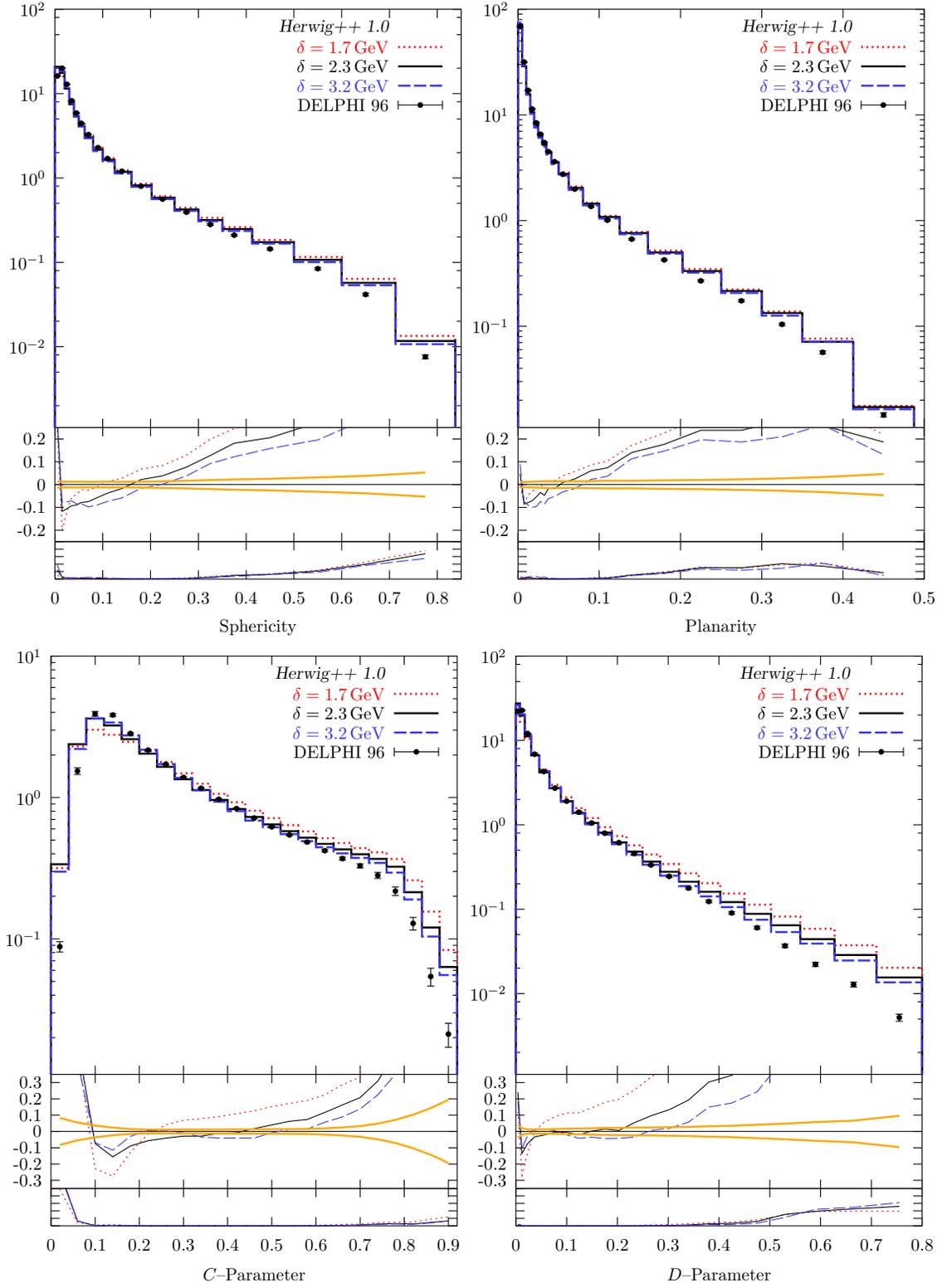
  \caption{Sphericity, planarity, $C$ parameter and $D$ parameter
  distributions. \label{fig:spcd}} 
}

\FIGURE[p]{
  \epsfig{file=es.25,scale=0.8}
  \epsfig{file=es.27,scale=0.8}\\[5pt]
  \epsfig{file=es.19,scale=0.8}
  \epsfig{file=es.21,scale=0.8}
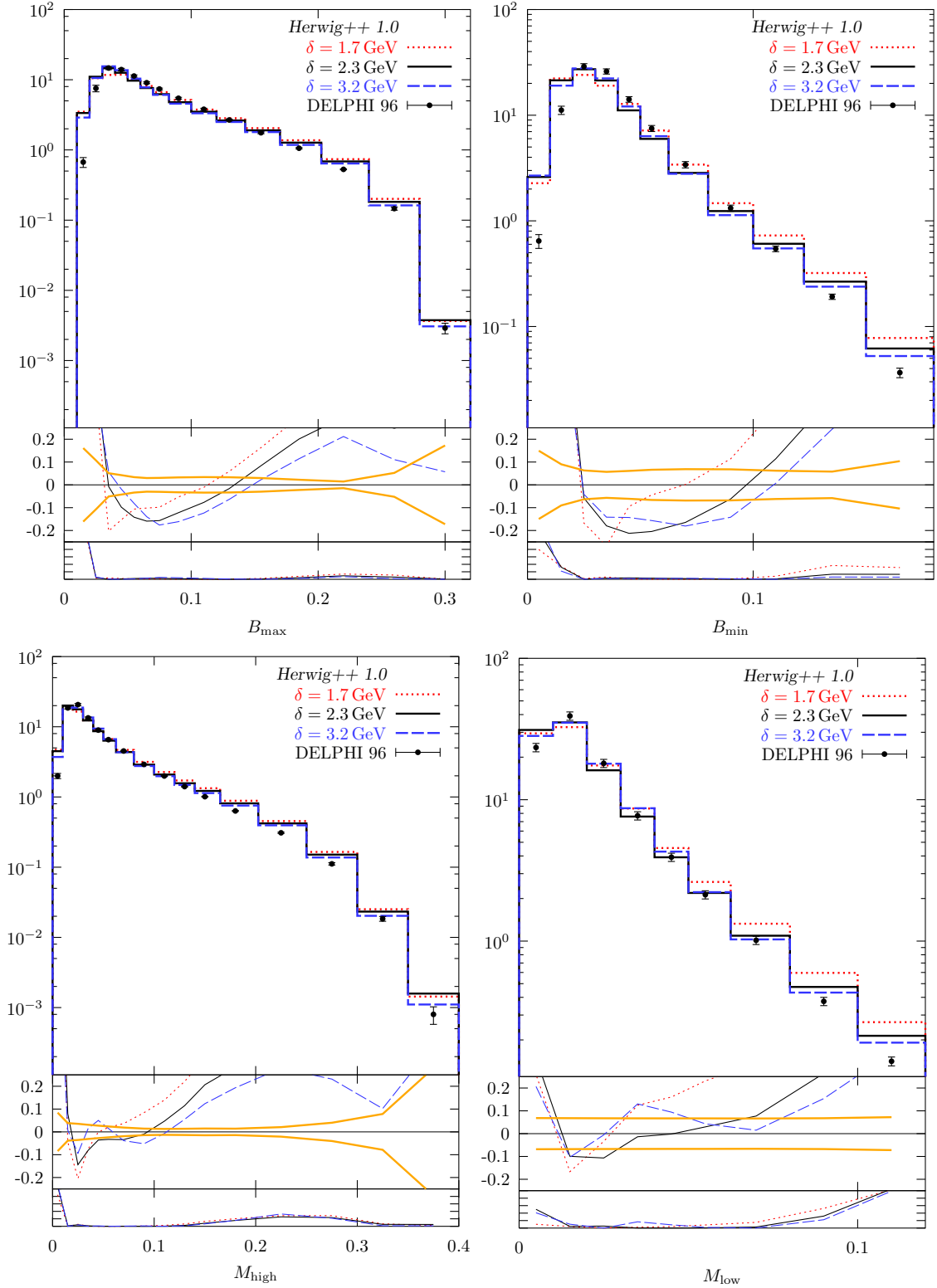
  \caption{The wide and narrow jet broadening measures
  $B_{\rm max}$ and $B_{\rm min}$ and the high and low hemisphere
  masses. 
  \label{fig:bmaxbmin}} 
}

\subsection{Four jet angles}
\label{sec:fourj}

We show the four-jet angles in fig.~\ref{fig:fourj}.  They
are considered only for events where we have a four-jet event at
$\ycut=0.008$.  Despite the fact that we do not have any matching to
higher order matrix elements, as was proposed in \cite{CKKW} and
implemented in \cite{Apacic}, the agreement between model and
data~\cite{DELPHIfourj} is remarkably good.  Even though we expected the
implementation of hard and soft matrix element corrections in \hpp{}
to be most important for the description of these observables, we did
not find very significant differences with or without the application
of matrix element corrections.

\FIGURE[p]{
  \epsfig{file=es.71,scale=0.8}
  \epsfig{file=es.73,scale=0.8}\\[5pt]
  \epsfig{file=es.75,scale=0.8}
  \epsfig{file=es.77,scale=0.8}
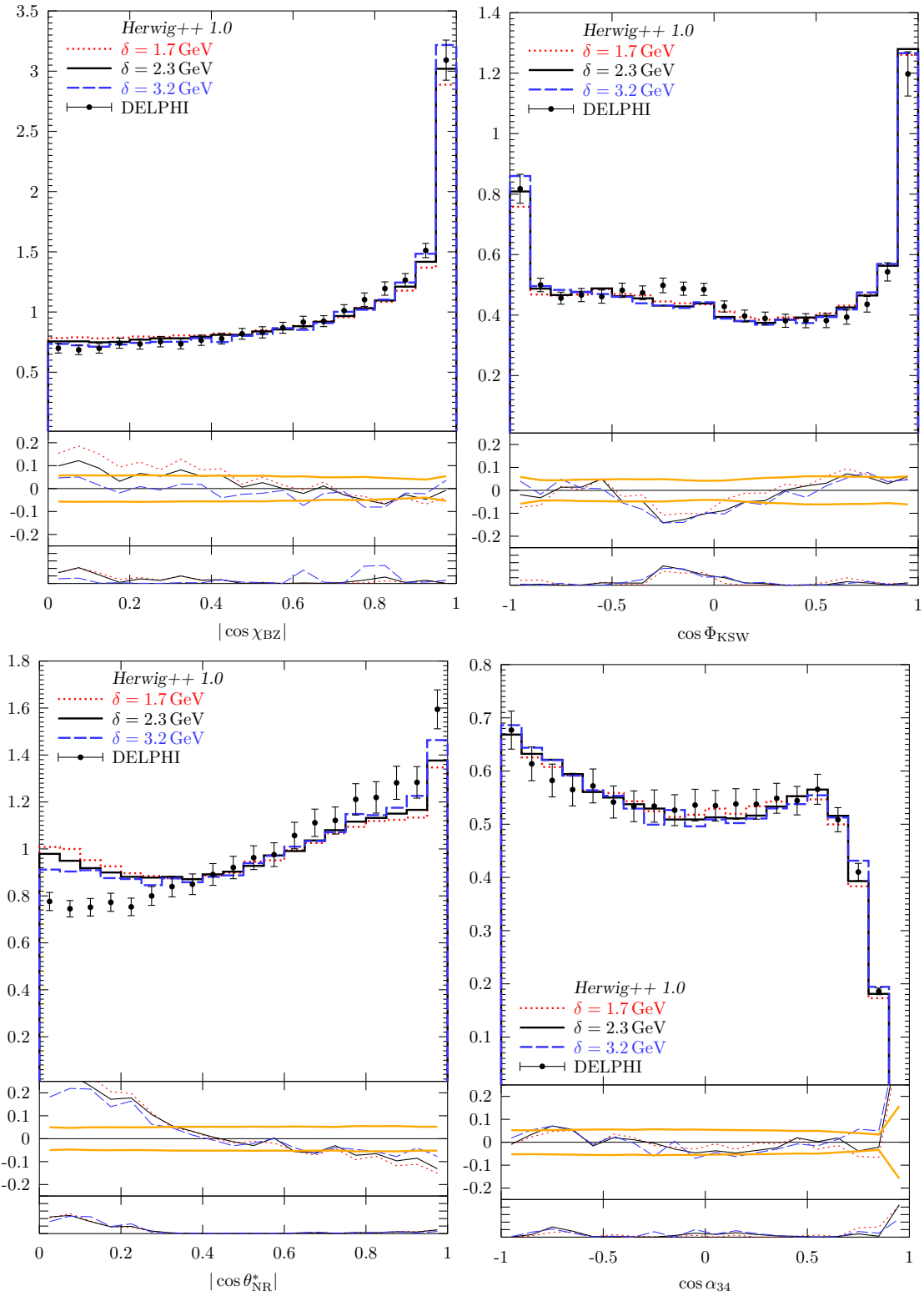
  \caption{Four jet angle distributions.  The points are
  from preliminary DELPHI data.\label{fig:fourj}}
}

\subsection{Single particle distributions}
\label{sec:singlep}

In fig.~\ref{fig:ptyT} we show single charged particle distributions
within the event, oriented along the thrust axis.  The transverse
momentum within the event plane $p_{\perp, {\rm in}}^T$ is shown with
and without matrix element corrections.  In contrast to the thrust
distribution we find that the matrix element corrections actually
improve the distribution.  Furthermore, $p_{\perp, {\rm out}}^T$ and
the rapidity along the thrust axis are rather well described.  We do
not show the analogous momentum distributions with respect to the
sphericity axis which have similar features.

We consider the distribution of scaled momentum $x_p=2|\bm p|/Q$ of
charged particles in fig.~\ref{fig:xpch}.  In addition to the full
distribution we also consider the results from light ($uds$), $c$ and
$b$ events.\footnote{The flavour of the quark-antiquark produced in
the initial hard process.}  In all cases we compare with data from
SLD \cite{SLD03}.
The charged particle distribution is well described in all
four cases, in fact somewhat better for heavy primary quarks.

\FIGURE[p]{
  \epsfig{file=es.32,scale=0.8}
  \epsfig{file=es.33,scale=0.8}\\[5pt]
  \epsfig{file=es.35,scale=0.8}
  \epsfig{file=es.37,scale=0.8}
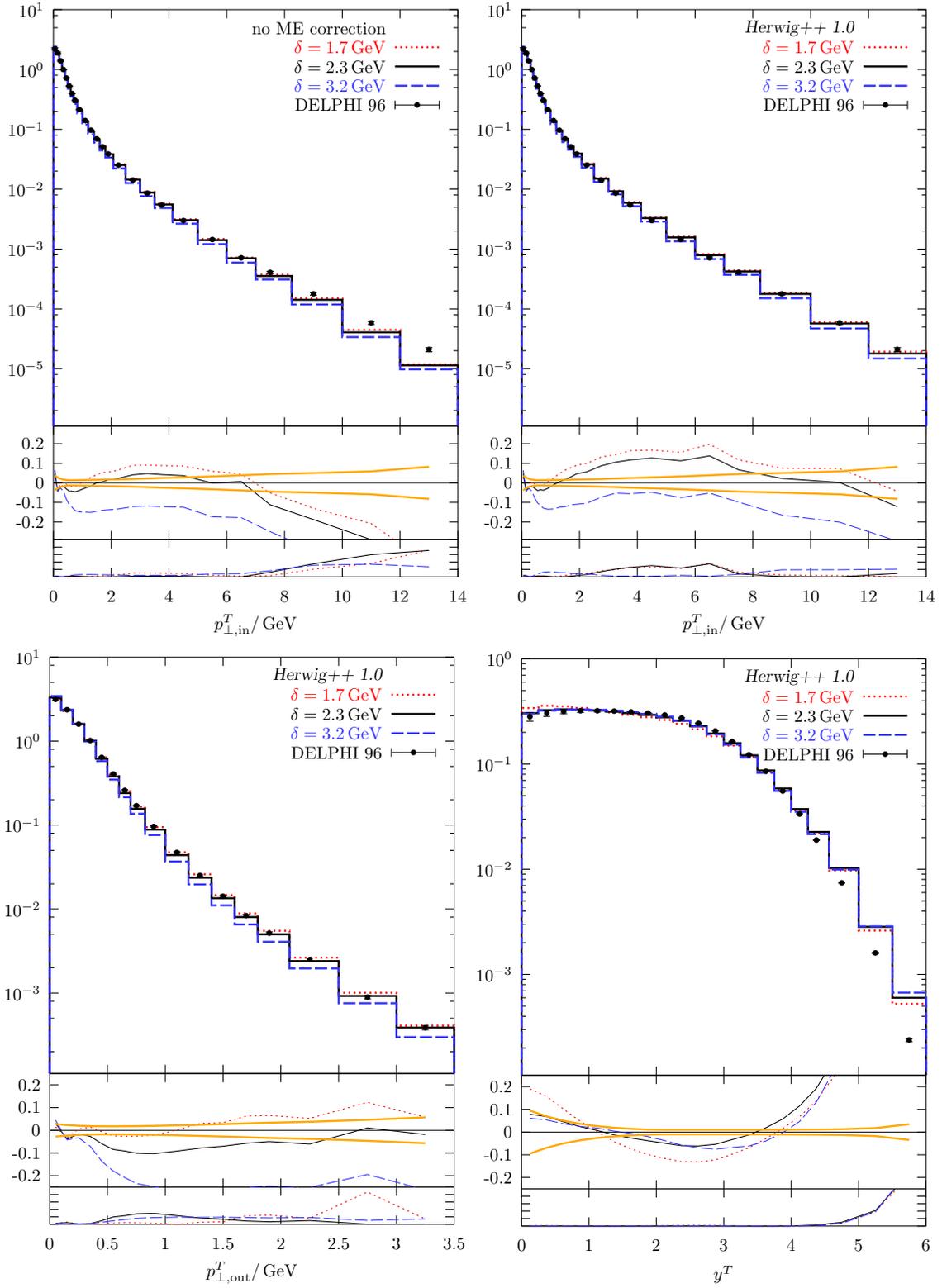
  \caption{Momentum distributions of charged particles with respect to
    the thrust axis, $p_{\perp, {\rm in}}^T$ (with and without matrix element
    corrections), $p_{\perp, {\rm out}}^T$ and $y^T$.
    \label{fig:ptyT}}
}

\FIGURE[p]{
  \epsfig{file=es.81,scale=0.8}
  \epsfig{file=es.83,scale=0.8}\\[5pt]
  \epsfig{file=es.85,scale=0.8}
  \epsfig{file=es.87,scale=0.8}
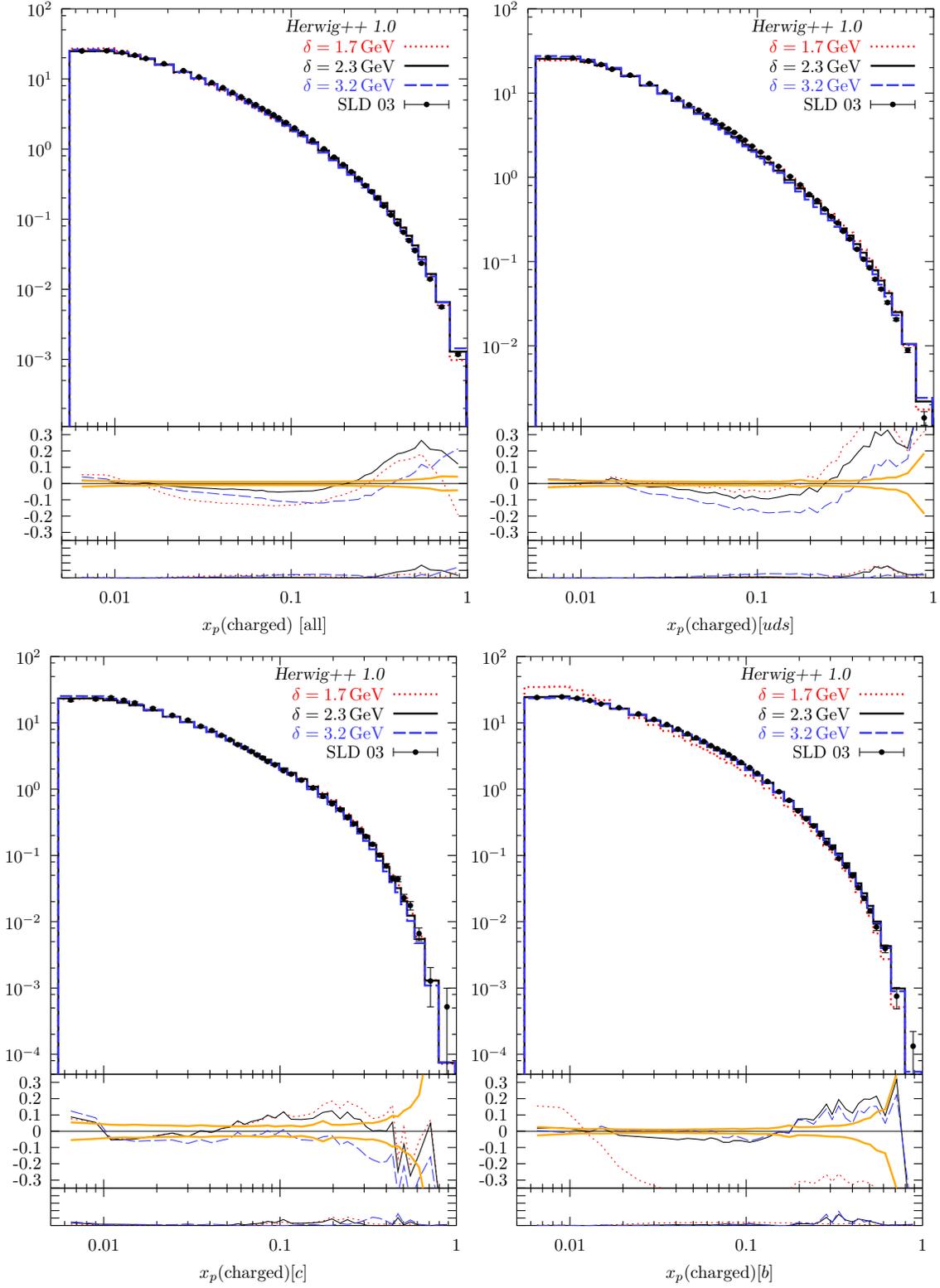
  \caption{The scaled momentum distribution $x_p$ of charged
    particles for all events as well as for $uds$,  $c$ and $b$
    events separately. \label{fig:xpch}} 
}

\subsection{Identified hadron spectra}
\label{sec:identhad}

As in the case of all charged particles we can compare identified
particle spectra from events of different flavour to SLD data
\cite{SLD03}.  Data for $\pi^\pm$ (not shown, being almost equivalent
to all charged particles), $K^\pm$ and ($p, \bar p$) are available.
In fig.~\ref{fig:xpp} we see the data for ($p, \bar p$) spectra from
events of different flavour.  For large values of $x_p$ we clearly
overshoot the data in light flavoured events.  This is somewhat
compensated by the heavy quark events which in turn seem to prefer
lower values of $x_p$.  We believe that this feature is related to the
hadronization, being similar to but smaller than that seen in \HW.

Fig.~\ref{fig:flavmom} shows distributions for $K^\pm$ and $\Lambda,
\bar \Lambda$.  Both are rather better described than the proton
spectra but the distribution of $\Lambda, \bar \Lambda$ tends to have
a similar, though smaller, `bump' in comparison to data from
ALEPH~\cite{ALEPHLambda}.

\FIGURE[p]{
  \epsfig{file=es.105,scale=0.8}
  \epsfig{file=es.107,scale=0.8}\\[5pt]
  \epsfig{file=es.109,scale=0.8}
  \epsfig{file=es.111,scale=0.8}
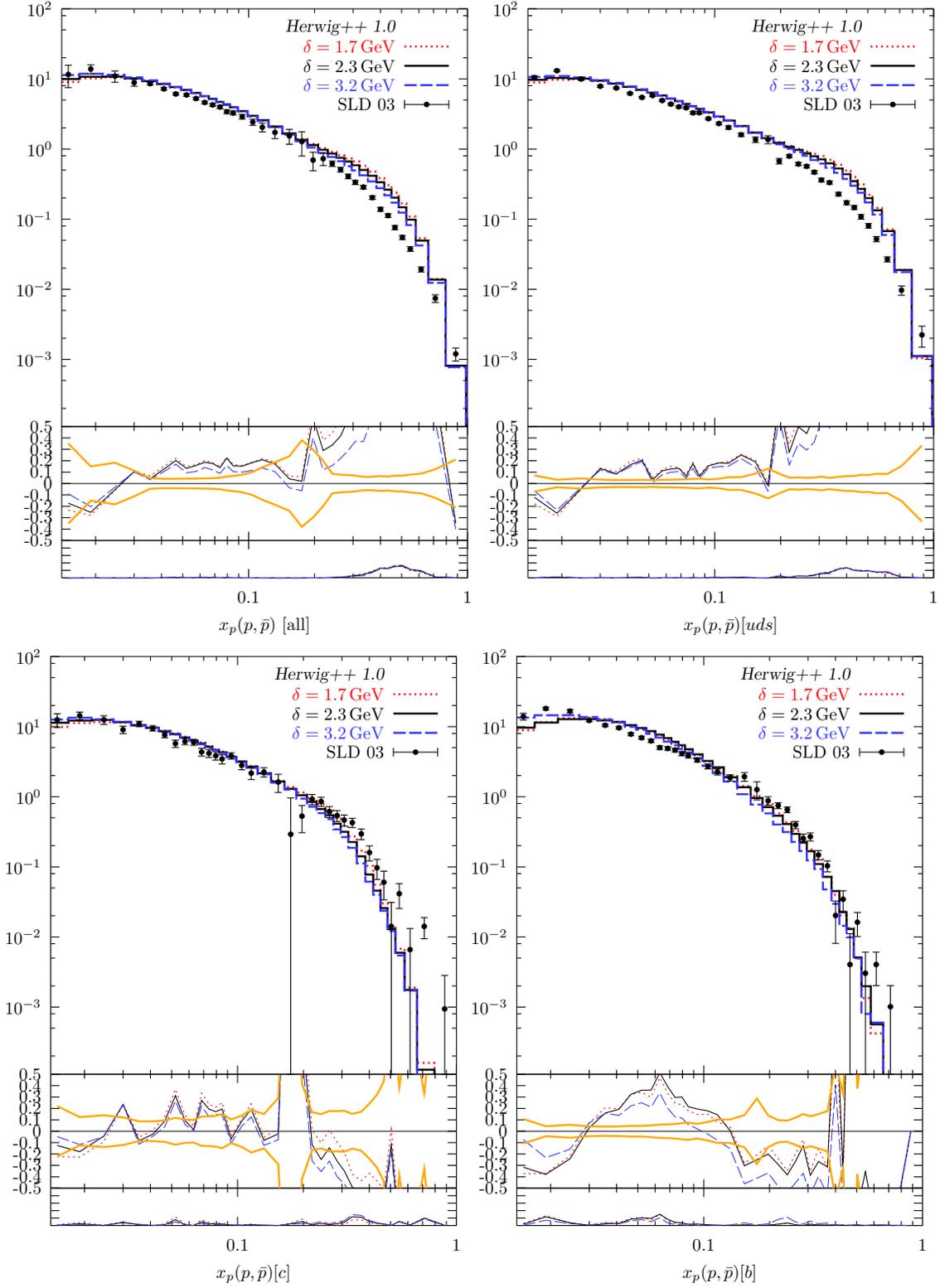
  \caption{The scaled momentum distribution $x_p$ of protons, shown
    separately for all events as well as for $uds$,  $c$ and $b$
    events. \label{fig:xpp}} 
}

\FIGURE[p]{
  \epsfig{file=es.97,scale=0.8}
  \epsfig{file=es.125,scale=0.8}
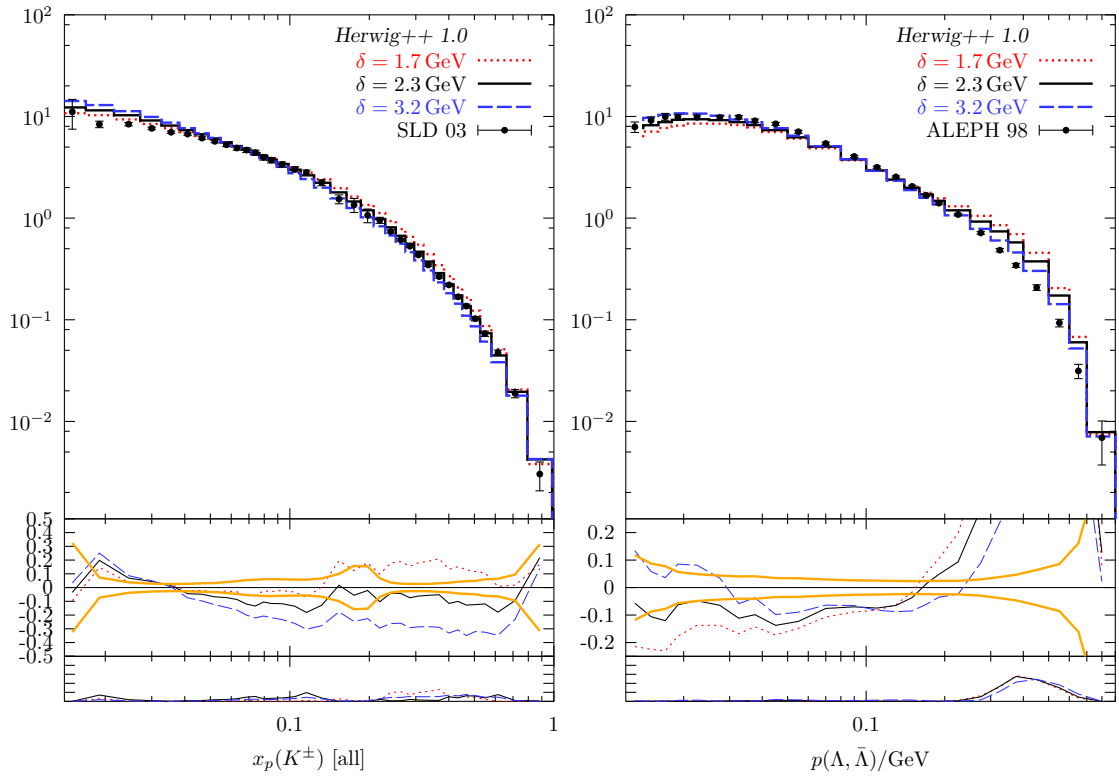
  \caption{Distribution of scaled kaon momentum and $\Lambda, 
    \bar \Lambda$ momentum. 
  \label{fig:flavmom}}
}

\subsection{B fragmentation function} 
In fig.~\ref{fig:bfraghad} we consider the $B$ hadron fragmentation
function in comparison to data from SLD \cite{SLDbfrag}.  We have also
considered data from ALEPH \cite{ALEPHbfrag} (not shown). We can
describe the data quite well without any additional tuning of the
hadronization model to this data.  The parton shower formulation in
terms of the new variables \cite{NewVariables} and taking quark masses
in the splitting functions into account clearly improves the
description of heavy quark events.

\FIGURE[p]{
  \epsfig{file=es.131,scale=.95}
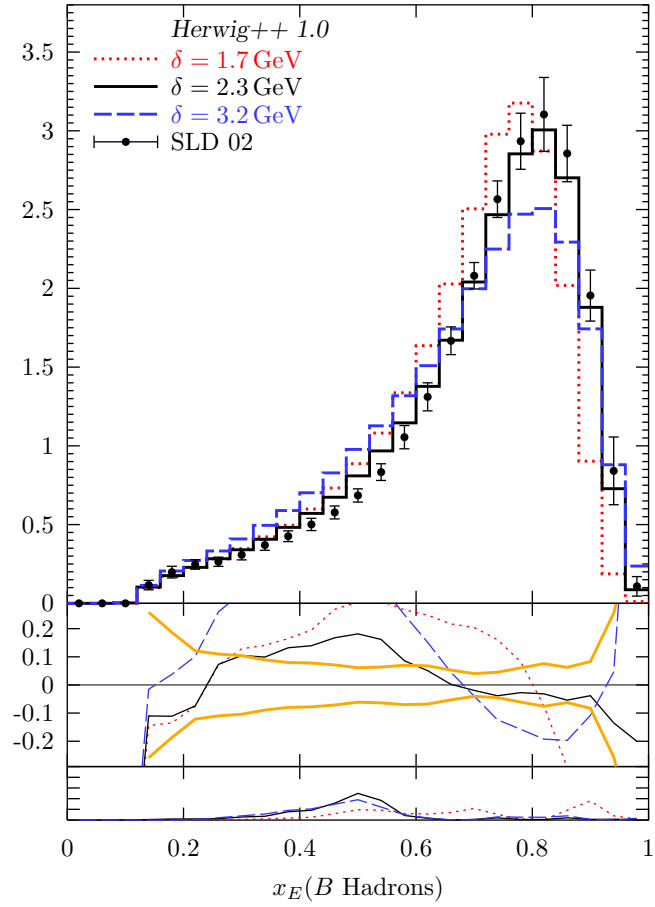
  \caption{The $B$--hadron fragmentation function for different
    values of the cutoff $\delta$.\label{fig:bfraghad}} }

\subsection{Overall results}

\TABLE{
\scriptsize
\begin{tabular}{lcrrrrrr}
\hline
\hline
&&\multicolumn{3}{c}{ME corrections off} 
&\multicolumn{3}{c}{ME corrections on}\\
Observable &Ref.& $\delta=1.7\,$GeV& $\delta=2.3\,$GeV &$\delta=3.2\,$GeV
& $\delta=1.7\,$GeV& $\delta=2.3\,$GeV &$\delta=3.2\,$GeV\\
\hline
\hline
$1-T$                      &\D    & 44.65& 33.15& 22.29& 72.80& 45.57& 26.34 \\
$M$                        &\D    &246.25&273.42&198.37&275.80&274.43&186.34 \\
$m$                        &\D    &150.74&157.91&137.43&174.29&163.02&129.10 \\
$O$                        &\D    &  7.41&  5.58&  5.14& 22.24& 19.33& 13.34 \\
\hline
$S$                        &\D    &  4.42&  3.50&  4.07& 24.70& 14.10&  8.50 \\
$P$                        &\D    &  4.48&  5.63&  6.54& 10.69&  7.40&  5.62 \\
$A$                        &\D    & 19.52& 10.80&  7.17& 44.83& 20.26& 11.33 \\
\hline
$C $                       &\D    & 66.86& 59.26& 39.56& 81.41& 67.44& 43.08 \\
$D $                       &\D    & 84.23& 29.30& 12.36&161.90& 60.42& 26.92 \\
\hline
$M_{\rm high}             $&\D    & 25.78& 18.88& 12.38& 38.43& 25.69& 11.52 \\
$M_{\rm low}              $&\D    & 15.25&  5.37&  2.50& 31.53& 10.42&  5.00 \\
$M_{\rm diff}             $&\D    &  7.28&  5.27&  7.25& 18.32& 12.17&  4.61 \\
\hline
$B_{\rm max}              $&\D    & 54.48& 50.29& 38.91& 59.61& 49.92& 33.23 \\
$B_{\rm min}              $&\D    & 53.25& 55.72& 53.18& 64.52& 58.08& 50.64 \\
$B_{\rm sum}              $&\D    &102.29& 97.35& 74.60&121.86&103.10& 70.98 \\
$B_{\rm diff}             $&\D    &  8.28&  5.42&  4.70& 18.39& 13.64&  6.09 \\
\hline
$p_{\perp, {\rm in}}^T    $&\D    &  2.48&  3.11& 11.52&  3.39&  1.70&  4.26 \\
$p_{\perp, {\rm out}}^T   $&\D    &  0.25&  3.28& 21.65&  0.80&  1.70& 16.06 \\
$y^T                      $&\D    & 34.52& 60.55& 66.05& 34.94& 53.81& 59.07 \\
\hline
$p_{\perp, {\rm in}}^S    $&\D    &  2.53&  3.19& 11.76&  2.32&  1.39&  4.30 \\
$p_{\perp, {\rm out}}^S   $&\D    &  0.37&  3.77& 22.64&  0.90&  2.01& 16.78 \\
$y^S                      $&\D    &  9.04& 17.49& 24.85&  7.78& 14.72& 21.94 \\
\hline
$D_2^D                    $&\D    &  9.37&  3.54&  3.76& 25.56& 11.27&  5.25 \\
$D_3^D                    $&\D    & 25.85&  6.33&  2.14& 47.11& 15.31&  5.42 \\
$D_4^D                    $&\D    & 43.90& 10.47&  2.69& 78.82& 23.26&  7.11 \\
\hline
$y_{23}                   $&\O    &  8.75&  6.11&  5.36& 12.35&  8.65&  6.40 \\
$y_{34}                   $&\O    & 10.20&  9.65&  9.07& 11.46& 10.02&  8.81 \\
$y_{45}                   $&\O    & 15.53& 14.40& 11.78& 17.74& 15.57& 11.75 \\
$y_{56}                   $&\O    & 16.02& 17.77& 15.13& 15.50& 17.51& 14.32 \\
\hline
$\langle N_{\rm jets}\rangle$&\O  & 12.84&  3.30&  0.62& 28.29& 12.80&  5.95 \\
\hline
$R_2                      $&\O    &  9.75&  6.56&  6.18& 19.84& 13.45&  9.59 \\
$R_3                      $&\O    & 10.46&  8.51&  9.36& 23.49& 15.86& 11.95 \\
$R_4                      $&\O    & 13.47& 10.95& 10.36& 15.26& 12.42& 10.22 \\
$R_5                      $&\O    & 25.53& 24.98& 23.43& 28.09& 26.35& 22.30 \\
$R_6                      $&\O    & 10.37&  1.74&  0.67& 18.38&  4.33&  1.47 \\
\hline
$\cos(\chi_{\rm BZ})      $&\DF   &  2.90&  1.10&  0.48&  2.48&  1.05&  0.53 \\
$\cos(\Phi_{\rm KSW})     $&\DF   &  2.30&  2.06&  2.56&  1.22&  1.50&  1.64 \\
$\cos(\theta_{\rm NR}^*)  $&\DF   &  7.68&  5.06&  2.72&  8.66&  6.22&  3.57 \\
$\cos(\alpha_{34})        $&\DF   &  1.41&  1.57&  1.71&  0.60&  0.64&  0.76 \\
\hline
$N_{ch}                   $&\OC   & 21.86& 25.71& 12.90& 19.81& 22.84& 12.97 \\
\hline
$x_p({\rm ch}) [\rm all]  $&\Sx   &  5.32&  5.65&  3.49&  4.75&  4.10&  3.02 \\
$x_p({\rm ch}) [uds]      $&\Sx   & 15.72&  8.50&  6.13& 12.63&  6.69&  5.86 \\
$x_p({\rm ch}) [c]        $&\Sx   &  3.95&  2.29&  2.17&  2.96&  1.76&  2.73 \\
$x_p({\rm ch}) [b]        $&\Sx   & 35.05&  3.23&  1.79& 35.79&  2.49&  1.22 \\
\hline
$x_p(\pi^\pm) [\rm all]   $&\Sx   &  8.29&  9.27&  6.18&  7.21&  7.50&  5.51 \\
$x_p(\pi^\pm) [uds]       $&\Sx   & 28.30& 15.92& 10.47& 24.05& 13.29&  9.46 \\
$x_p(\pi^\pm) [c]         $&\Sx   &  4.65&  2.99&  1.38&  3.67&  2.28&  1.62 \\
$x_p(\pi^\pm) [b]         $&\Sx   & 49.13&  3.14&  1.56& 49.44&  3.57&  2.02 \\
\hline
$x_p(K^\pm) [\rm all]     $&\Sx   &  4.99&  2.02& 15.38&  3.67&  2.88& 17.37 \\
$x_p(K^\pm) [uds]         $&\Sx   &  6.46& 17.05& 36.45&  6.83& 19.36& 38.79 \\
$x_p(K^\pm) [c]           $&\Sx   & 21.01&  2.20&  3.35& 18.16&  1.75&  4.14 \\
$x_p(K^\pm) [b]           $&\Sx   &  8.56&  7.14&  4.34&  7.63&  5.84&  4.97 \\
\hline
$x_p(p, \bar p) [\rm all]$ &\Sx   &143.34& 98.19& 42.90&140.48& 87.08& 36.23 \\
$x_p(p, \bar p) [uds]$     &\Sx   &145.35&102.51& 52.78&139.85& 91.07& 45.10 \\
$x_p(p, \bar p) [c]$       &\Sx   &  2.26&  2.41&  2.86&  2.34&  2.48&  2.85 \\
$x_p(p, \bar p) [b]$       &\Sx   & 11.26& 13.71&  8.12& 11.47& 13.54&  8.31 \\
\hline
$p(\Lambda, \bar \Lambda)$ &\A    & 58.02& 28.52&  9.47& 55.27& 25.50&  7.86 \\
\hline
$x_E (B)$                  &\SB   &  8.93&  0.92&  8.16&  9.44&  1.39&  9.92 \\
$x_E (B)$                  &\AB   & 15.40&  1.75&  7.35& 15.76&  2.01&  8.21 \\
\hline
\hline
$\langle\chi^2\rangle/{\rm bin}$&& 32.75& 25.84& 20.93& 40.69& 28.41& 20.56 \\
\hline
\hline
\end{tabular}\centering
\caption{$\chi^2/$bin for all observables we studied and a relevant subset of
  parameters. \label{tab:chi2}}}

In table~\ref{tab:chi2} we show a list of $\chi^2$ values for all
observables that were studied during our analysis, including those
not shown in the plots.  The most sensitive parameters were the
cutoff value $\delta$ and the use of (hard plus soft) matrix element
corrections.  The table shows three values of $\delta$:  our preferred
value of $\delta=2.3\,$GeV as well as the lowest and highest values
that we considered.  

The results should be interpreted with care.  The overall trend
suggests that we should prefer a large cutoff scale.  However, we
have just averaged over all possible observables.  Taking a closer
look, we may want to weight different observables in a different way.  

In more detail, the general trend is the following: event shapes,
jet rates and differential jet rates prefer a low cutoff. The single
particle distributions along the thrust and sphericity axes prefer a
small cutoff value. The $y_{nm}$ distributions prefer either a high or
a low cutoff value.  The spectra of identified particles tend to
prefer the high cutoff value with some exceptions for light quark
events.  The $B$ fragmentation function clearly prefers the
intermediate value.  

In addition, as indicated in sec.~\ref{sec:multcomment}, we found
that the measured yields of identified particles clearly prefer the
value $\delta=2.3\,$GeV.

\section{Conclusions}
\label{sec:conclusions}
We have achieved a complete event generator for $\ee$ annihilation into
hadrons.  The main physics features, in comparison to the previous
versions of \HW, are an
improved parton shower, capable of properly describing the
perturbative splitting of heavy quarks, and an improved cluster
hadronization model.

We have tested our model against a wide range of data from $\ee$
colliders and are able to give a good general description of the data.

For many observables the description of the data has been improved
with respect to \HW{}.  The new parton shower has a number of
remarkable features. The need for matrix element corrections has
decreased.  The main reason for this is the use of improved
splitting functions, which give a far better approximation of the
matrix elements in the region of collinear gluon emissions.  We can
describe observables involving light or heavy quark splitting with a
unique set of parameters. The new hadronization model also improves
the description of identified particle spectra and multiplicities.

The detailed analysis of
our results leaves us with a \emph{recommendation}: the set of
parameters that is shown in table~\ref{tab:Herwig_defaults}.  This set
of parameters is understood as a weighted compromise in order give a
good overall description of the data we have considered so far.
We did not aim at a complete tuning of the model, but rather wanted
to study its ability to  describe the broad features
of the data, which turned out to be very successful.

Future work on the program will extend the parton shower to initial
state radiation and include a model for the soft underlying event in
hadron--hadron collisions, aiming at a complete event generator for
the simulation of Tevatron and LHC events.

\acknowledgments 
We would like to thank Leif L\"onnblad for constantly
supporting \ThePEG{} and for many useful conversations.  We also thank
Peter Richardson for very fruitful discussions and comments.  Many
thanks to Klaus Hamacher, Hendrik Hoeth and G\"unther Dissertori for
providing us with four-jet data.  S.G.\ and P.S.\ would like to thank
the CERN Theory Division for hospitality while part of this work was
done.  S.G.\ would like to thank Frank Krauss and David Ward for
fruitful conversations. This work was supported by the UK Particle
Physics and Astronomy Research Council.

\end{document}